%% file: root.tex
\documentclass[twocolumn,letterpaper]{IEEEtran}

\input{packages}
\input{macros}

\usepackage{ifthen}
\newboolean{arxiv}
\setboolean{arxiv}{true}
\newcommand{\IfArxiv}[2]{\ifthenelse{\boolean{arxiv}}{{\color{teal}#1}}{{#2}}}
\newcounter{defcounter}
\newenvironment{xequation}{%
\addtocounter{equation}{-1}
\refstepcounter{defcounter}
\renewcommand\theequation{X\thedefcounter}
\begin{equation}}
{\end{equation}}

\usepackage{geometry}
\title{Robust Synergistic Hybrid Feedback\IfArxiv{\\(Extended Version)}{}}
\author{Pedro Casau, Ricardo G. Sanfelice,~\IEEEmembership{Fellow, IEEE,} and Carlos Silvestre~\IEEEmembership{Senior Member, IEEE,}%
\thanks{P. Casau is with the Department of Electrical and Computer Engineering at Instituto Superior Técnico, Universidade de Lisboa, Lisboa, Portugal. E-mail address: \texttt{pcasau@isr.tecnico.ulisboa.pt}. %
C. Silvestre is with the Department of Electrical and Computer Engineering of the Faculty of Science and Technology of the University of Macau, Macau, China, and with Instituto Superior Técnico, Universidade de Lisboa, Lisboa, Portugal. E-mail address: \texttt{csilvestre@um.edu.mo}. %
R. G. Sanfelice is with the Department of Computer Engineering, University of California, Santa Cruz, CA 95064. Email address: \texttt{ricardo@ucsc.edu}. %
This work was partially supported by the Macao Science and Technology Development Fund under Grant FDCT/0146/2019/A3, by the University of Macau, Macao, China, under Project MYRG2020-00188-FST, by the Fundação para a Ciência e a Tecnologia (FCT)  through LARSyS - FCT Project UIDB/50009/2020 and LAETA - FCT Project UIDB/50022/2020, and by FCT Scientific Employment Stimulus grant CEECIND/04652/2017. Research by R. G. Sanfelice has been partially supported by the National Science Foundation under Grant no. ECS-1710621, Grant no. CNS-1544396, and Grant no. CNS-2039054, by the Army Research Office under Grant no. W911NF-20-1-0253, and by the Air Force Office of Scientific Research under Grant no. FA9550-19-1-0053, Grant no. FA9550-19-1-0169, and Grant no. FA9550-20-1-0238.}%
}

\date{}

\newcommand{\T}[1][]{\mathsf{T}_{#1}}
\newcommand{\tto}{\rightrightarrows}
\newcommand{\clarke}[1]{#1^{\,\circ}}
\renewcommand{\bar}[1]{\overline{#1}}
\DeclareMathOperator{\gph}{gph}
\DeclareMathOperator{\Proj}{Proj}
\DeclareMathOperator{\p}{p}
\newcommand{\KL}{\mathcal{KL}}
\renewcommand{\bd}{\partial}
\renewcommand{\cl}[1]{\bar{#1}}
\newcommand{\radius}{r}

\newcommand{\F}[1][]{F_{#1}}

\newcommand{\ee}[1][]{e_{#1}}

\renewcommand{\Amc}[1][]{\mathcal{A}_{#1}}
\newcommand{\Xmc}[1][]{\mathcal{X}_{#1}}

\newcommand{\Umc}[1][]{\mathcal{U}_{#1}}
\newcommand{\Hmc}[1][]{\mathcal{H}_{#1}}
\newcommand{\Bmc}{\mathcal{B}}
\newcommand{\Emc}[1][]{\mathcal{E}_{#1}}

\newcommand{\Nmc}{\mathcal{N}}

\newcommand{\Ascr}[1][]{\mathscr{A}_{#1}}
\newcommand{\Vscr}[1][]{\mathscr{V}_{#1}}

\newcommand{\xc}[1][]{%
	\IfEqCase{#1}{%
        {2}{x_{c,2}}%
		{1}{x_{c,1}}%
		{0}{x_{c,0}}%
	}[x_c]%
}
\let\olddot\dot
\renewcommand{\dot}[1]{%
	\IfEqCase{#1}{%
		{xc}{\olddot{x}_{c}}%
		{xc0}{\olddot{x}_{c,0}}%
		{xc1}{\olddot{x}_{c,1}}%
		{xc2}{\olddot{x}_{c,2}}%
	}[\olddot{#1}]%
}
\newcommand{\V}[1][]{V_{#1}}
\newcommand{\minV}[1]{\nu_{#1}}
\newcommand{\qminV}[1]{\varrho_{#1}}
\newcommand{\jset}[1][]{D_{#1}}
\newcommand{\jmap}[1][]{G_{#1}}
\newcommand{\fset}[1][]{C_{#1}}
\newcommand{\fmap}[1][]{F_{#1}}
\newcommand{\gap}[1]{\mu_{#1}}

\let\oldpi\pi
\renewcommand{\pi}[1][]{\oldpi_{#1}}
\let\oldpsi\psi
\renewcommand{\psi}[1][]{%
    \IfEqCase{#1}{%
        {x}{f}%
        {u}{H}%
        {theta}{W}%
        {th}{\widehat{W}}%
    }[\oldpsi_{#1}]%
}
\let\oldphi\phi
\renewcommand{\phi}[1][]{\oldphi_{#1}}
\let\oldkappa\kappa
\renewcommand{\kappa}[1][]{\oldkappa_{#1}}
\let\oldgamma\Gamma
\renewcommand{\Gamma}[1][]{\oldgamma_{#1}}
\newcommand{\uc}[1][]{%
	\IfEqCase{#1}{%
		{theta}{u_{c,\theta}}%
	}[u_c]%
}
\newcommand{\ud}[1][]{%
	\IfEqCase{#1}{%
		{theta}{u_{d,\theta}}%
	}[u_d]%
}
\newcommand{\deltabnd}[1][]{\bar{\delta}_{#1}}
\newcommand{\deltalow}{\underline{\delta}}

\renewcommand{\th}{\hat{\theta}}
\newcommand{\slack}{\epsilon}
\newcommand{\tz}{\theta_0}
\renewcommand{\tt}{\tilde{\theta}}
\let\oldPsi\Psi
\renewcommand{\Psi}[1][]{\oldPsi_{#1}}
\newcommand{\ku}{k_u}
\newcommand{\ff}{\upsilon}

\newcommand{\U}[1][]{U_{#1}}

\newcounter{asscnt}
\newcounter{deltacnt}

\PassOptionsToPackage{normalem}{ulem}
\usepackage{ulem}

\newcommand{\add}[1]{{\color{red}{}#1}}

\usepackage{fancyhdr}

\begin{document}
\ifthenelse{\boolean{arxiv}}{%
\pagestyle{fancy}
\fancyhead[C]{Submitted for publication}
}{}
\maketitle
\ifthenelse{\boolean{arxiv}}{\vspace*{-15pt}}{}
\begin{abstract}
Synergistic hybrid feedback refers to a collection of feedback laws that allow for global asymptotic stabilization of a compact set through the following switching logic: given a collection of Lyapunov functions that are indexed by a logic variable, whenever the currently selected Lyapunov function exceeds the value of another function in the collection by a given margin, then a switch to the corresponding feedback law is triggered. This kind of feedback has been under development over the past decade and it has led to multiple solutions for global asymptotic stabilization on compact manifolds. 
The contributions of this paper include
a synergistic controller design in which the logic variable is not necessarily constant between jumps,
a synergistic hybrid feedback that is able to tackle the presence of parametric uncertainty,
backstepping of adaptive synergistic hybrid feedbacks, and a
demonstration of the proposed solutions to the problem of global obstacle avoidance.
\end{abstract}
\section{Introduction}
\subsection{Background and Motivation}


In this paper, we consider the problem of globally asymptotically stabilizing continuous-time plants of the form
\begin{equation}\label{eq:plant}
    \dot{x}_p=F_p(x_p,u_p,\theta)
\end{equation}
where $x_p\in\Xmc[p]$ denotes the state of the plant, $u_p\in\Umc[p]$ is the input, and $\theta$ represents the parameters of the plant. To this end, we propose the following hybrid controller 
\begin{equation}\label{eq:contrived}
\begin{aligned}
    \begin{drcases}
        \dot{\chi}_c\in\hat{\fmap[c]}(x_p,\chi_c,\xc,u_c)\\
        \dot{xc}\in\fmap[c](x_p,\chi_c,\xc)
    \end{drcases}& & &(x_p,\chi_c,\xc)\in\fset,\ u_c\in\Umc[c]\\
    \begin{drcases}
        \chi_c\pl=\chi_c\\
        \xc\pl\in\jmap[c](x_p,\chi_c,\xc)
    \end{drcases}& & &(x_p,\chi_c,\xc)\in\jset
\end{aligned}
\end{equation}
where $\chi_c\in\hat{\Xmc[c]}$ and $\xc\in\Xmc[c]$ represent different components of the state of the controller, $\hat{\fmap[c]}$ and $\fmap[c]$ are the flow maps associated with $\chi_c$ and $\xc$, respectively, $\fset$ denotes the flow set, $\jmap[c]$ defines the update law for jumps of $\xc$ and $\jset$ is the jump set. The key differences between $\chi_c$ and $\xc$ is the fact that $\chi_c$ does not change its value during jumps and also that the flows of $\chi_c$ depend on a virtual input variable $u_c\in\Umc[c]$. 
More precisely, the goal in this paper is to design a controller that globally asymptotically stabilizes a compact set $\Amc$ for the closed-loop system resulting from the interconnection between~\eqref{eq:plant} and~\eqref{eq:contrived} both when the parameter $\theta$ is known, but also when it is only known to belong to a given compact set ~$\Omega$.

In the presence of topological obstructions, this objective is not attainable via continuous feedback and, even though it might be attainable through discontinuous feedback, the resulting closed-loop system may not be robust to arbitrarily small noise (cf.~\cite{bhat_topological_2000} and~\cite{mayhew_topological_2011}). 
To illustrate these limitations of continuous/discontinuous feedback, let us consider the problem of globally asymptotically stabilizing the point $(1,0)$ for the dynamical system
\begin{align*}
\dot{x}_1&=-x_2 u_p,&
\dot{x}_2&=x_1 u_p,
\end{align*}
where $x_p\ceq(x_1,x_2)\in{\Xmc\ceq}\sphere 1\ceq\{x_p\in\R{2}:\norm{x_p}=1\}$ is the state variable and $u_p\in\R{}$ denotes the input. In this direction, let 
$h(x_p)=(1-x_1)/2$
for each $x_p\in\sphere 1$. The gradient-based feedback law is given by 
$u_p= \bmtx{x_2& -x_1}\grad h(x_p)$
which represents the projection of the gradient of $h$ onto the tangent space to $\sphere 1$ at $x_p$. It follows from standard Lyapunov stability arguments that $(1,0)$ is asymptotically stable for the closed-loop system, but it is not globally asymptotically stable since $x=(-1,0)$ is also an equilibrium point.

It can be argued that the discontinuous feedback law%
\begin{equation}
\label{eq:discontinuous}
u_p=\kappa[p](x_p)=\begin{cases}
-1 & \text{ if } x_1 = -1\\
0 & \text{ if } x_1 = 1\\
-\frac{\displaystyle x_2}{\displaystyle \norm{x_2}} & \text{ otherwise} 
\end{cases}
\end{equation}
defined for each $x_p\in\sphere 1$ globally asymptotically stabilizes $(1,0)$ if one considers Carathéodory solutions to the discontinuous closed-loop system because, in this case, $(-1,0)$ is not an equilibrium point. However, due to the discontinuity of the feedback law~\eqref{eq:discontinuous}, arbitrarily small noise can induce chattering which is a property that is ellucidated by considering generalized solutions to discontinuous dynamical systems such as Krasovskii solutions (cf.~\cite{mayhew_quaternion-based_2011}).
These limitations of continuous and discontinuous feedbacks constitute the motivation for the development of synergistic hybrid feedback.

If $\hat{\fmap[c]}$ in~\eqref{eq:contrived} defining the dynamics of $\chi_c$ is given, then $\chi_c$ can become part of the state of~\eqref{eq:plant} and the stated objective can be attained through the design of a hybrid controller $\Hmc[c]\ceq(\fset,\fmap[c],\jset,\jmap[c])$ with state $\xc\in\Xmc[c]$ and dynamics
\begin{subequations}
\begin{align*}
    \dot{xc}&\in\fmap[c](x,\xc) & (x,\xc)&\in\fset\\
    \xc\pl&\in\jmap[c](x,\xc) & (x,\xc)&\in\jset
\end{align*}
\end{subequations}
assigning $u\ceq(u_p,u_c)\in\Umc\ceq\Umc[p]\x\Umc[c]$ via a feedback law $(x,\xc)\mapsto\kappa(x,\xc)$, where $x\ceq(x_p,\chi_c)\in\Xmc\ceq\Xmc[p]\x\hat{\Xmc[c]}$ is the state of the system to control with dynamics described by the following differential inclusion 
\begin{align}
    \label{eq:sysx}\dot{x}\in \fmap[\theta](x,\xc,u)\ceq\fmap[p](x_p,u_p,\theta)\x\hat{\fmap[c]}(x_p,\chi_c,\xc,u_c),
\end{align}
where $\theta$ is a constant.

This formulation enables the controller design for systems whose dynamics depend on the controller state. For example, given a plant with dynamics
$\dot{x}_p=f_p(x_p)+H_p(x_p)u_p+W_p(x_p)\theta$
where $f_p,H_p, W_p$ are functions with the appropriate dimensions, suppose that the reference trajectory to be tracked is denoted by $x_d$ and that it is generated by the system 
$\dot{x}_d=f_p(x_d)+H_p(x_d)\xi_d$
for some signal $\xi_d$. The tracking error $x\ceq x_p-x_d$ can be taken as the state of the system~\eqref{eq:sysx}, in which case we have that
$\fmap[\theta](x,\xc,u)=f_p(x_d+x)-f_p(x_d)-H_p(x_d)\xi_d
+H_p(x_d+x)u+W_p(x_d+x)\theta$
by identifying $u_p$ with $u$ and by considering $(x_d,\xi_d)$ as components of the controller variable $\xc$. More practically, $x$ can be considered to be the part of the state of the closed-loop system that remains unchanged during jumps.
In this paper, we present two novel synergistic hybrid controllers for global asymptotic stabilization of a compact set for a closed-loop system with the plant dynamics in~\eqref{eq:sysx}. The first controller design considers that the parameter $\theta$ is known, while the second controller design considers that $\theta$ is unknown but belongs to a known compact set $\Omega$.

\subsection{Literature Review}

Synergistic hybrid feedback is a hybrid control strategy that consists of a collection of potential functions that asymptotically stabilize a given compact set by gradient descent{ feedback}. If, for all equilibria that do not lie within the given compact set, there exists another function in the collection {that} has a lower value and does not share the same equilibria, then it is possible to achieve global asymptotic stabilization of the given compact set through hysteretic switching {(see, e.g.,~\cite{Mayhew2011})}. 

Synergistic hybrid feedback came to prominence with the work~\cite{mayhew_quaternion-based_2011} on quaternion-based feedback for global asymptotic stabilization attitude tracking, thereby solving the attitude control problem (cf.~\cite{wen_attitude_1991}).
The framework of synergistic hybrid feedback provides not only a solution to the problem of attitude control but, more importantly, it provides {a} robust solution for global asymptotic stabilization on compact manifolds. The works~\cite{Nakamura2019} and~\cite{Casau2019} leverage the concepts at the root of synergistic hybrid feedback and use them to design controllers that are applicable to a broad class of systems. 
However, most of the contributions on this class of hybrid controllers are on the control of robotic systems, such as pendulum stabilization~\cite{mayhew_global_2010}, vector-based rigid body stabilization~\cite{berkane_construction_tac2016,Casau2015b}, tracking for marine and aerial vehicle~\cite{casau_robust_2015,Basso2022}, {and }rigid body tracking through rotation matrix feedback~\cite{mayhew_synergistic_tac2013,lee_global_tac2015}. 
Within the field of robotics, we single out the problem of obstacle avoidance, which is also addressed in this paper. 

Obstacle avoidance is an important and longstanding problem that reflects the need to drive a {the state of }a system from one place to another while avoiding obstacles in its way. Several solutions to this problem have been proposed over the last few decades as highlighted in~\cite{Rybus2018}. In particular, it is possible to find both stochastic~\cite{malone_hybrid_2017} as well as deterministic approaches~\cite{Bloch2021} to tackle the obstacle avoidance problem.
However, it was shown in~\cite{Kod1990} that in a ``sphere world,'' there is at least one saddle equilibrium point for each obstacle within the state space, thus precluding global asymptotic stabilization of a setpoint by continuous feedback. To address this limitation, hybrid control solutions to the problem of obstacle avoidance were proposed in~\cite{Sanfelice2006},~\cite{berkane_hybrid_2019},~\cite{Casau2019} {and~\cite{Marley2021}}.

Though not directly addressed in this paper, the concepts of synergistic hybrid feedback have also been used for observer design, optimization and control barrier function design in in~\cite{berkane_observer_2017},~\cite{strizic_hybrid_2017}, and~\cite{Marley2021}, respectively.

\subsection{Contributions}\label{sec:contributions}
The contributions in this paper are as follows:
\begin{inparaenum}
\item We develop a {dynamic synergistic hybrid feedback controller for global asymptotic stabilization of a broad class of dynamical systems}. In particular, we consider that the distinguishing feature of synergistic hybrid feedback is the switching logic, thus we depart from earlier works which were limited to controller variables that were constant during flows;
\item We provide a modification to the {dynamic} synergistic controller that takes into account the presence of {parametric uncertainty};
\item We demonstrate how the proposed constructions can be used to develop an adaptive synergistic controller for the stabilization of compact sets for affine control systems under matched uncertainties;
\item We show that the proposed adaptive controller is amenable to hybrid backstepping;
\item We apply the proposed controller designs to the problem of global obstacle avoidance in the presence of {parametric uncertainty} and illustrate the behavior of the closed-loop system through simulations.
\end{inparaenum}

The paper is organized as follows: in Section~\ref{sec:setup} we present the main assumptions on the plant dynamics{. In} Section~\ref{sec:basic} we provide the conditions under which the closed-loop system is well-posed{. In} Section~\ref{sec:gas} we provide sufficient conditions for global asymptotic stability of a compact set for the closed-loop system{. In} Section~\ref{sec:robust}, we develop the concept of robust synergistic hybrid feedback{. In} Section~\ref{sec:adaptive}, we apply the synergistic approach to the development of an adaptive synergistic controller for stabilization of affine control systems subject to matched uncertainties{. In} Section~\ref{sec:obstacles}, we apply the proposed controller to the problem of global obstacle avoidance{. In} Section~\ref{sec:conclusion}, we present some concluding remarks. 

A preliminary version of this paper was presented at the 2019 ACC with a simpler synergistic controller design for global asymptotic stabilization of control affine systems and without the full proofs (cf.~\cite{Casau2019a}). \IfArxiv{The original version of this paper has been submitted for publication.}{An extended version of this paper can be found at~\cite{casau_arxiv_2022}.}

\section{Notation \& Preliminaries}\label{sec:not}
\input{notation.tex}

\section{Problem Setup}\label{sec:setup}
Given sets $\Xmc$, $\Xmc[c]$, and $\Umc$, we consider a dynamical system with state $x\in\Xmc$ that is governed by the dynamics~\eqref{eq:sysx}
where $\xc\in\Xmc[c]$ is a controller variable, $u\in\Umc$ is the input, $\theta$ is a constant parameter that belongs to a compact set $\Omega$ and $\F[\theta]$ is a \hyperlink{def:setvaluedmap}{set-valued map} with the following properties.
\begin{assumption}\label{ass:data}
    Given sets $\Xmc$, $\Xmc[c]$, $\Umc$, and $\F[\theta]$ as in~\eqref{eq:sysx} the following properties hold:
    \begin{enumerate}[label=(S\arabic*)]
    \item \label{ass:star} Each set $\Xmc$, $\Xmc[c]$ and $\Umc$ is a closed nonempty subset of some \hyperlink{def:Euclid}{Euclidean space};
    \item\label{ass:Fbasic}{The set-valued map $\F[\theta]$} is outer semicontinuous, locally bounded, and convex-valued.
    \end{enumerate}
\end{assumption}
{Assumption~\ref{ass:star}} allows for the use of the analysis tools for hybrid dynamical systems that are provided in~\cite{goebel_hybrid_2012} which consider sets as subspaces 
of Euclidean spaces with the Euclidean metric topology. 
Assumptions~\ref{ass:VC} and~\ref{ass:Fbasic} are used to prove that the resulting closed-loop system has nontrivial solutions and that it satisfies the \hyperlink{def:hbc}{hybrid basic conditions}, respectively.

\begin{remark}\label{rem:hausdorff}
Since the sets $\Xmc$, $\Xmc[c]$ and $\Umc$ are closed relative to their Euclidean ambient spaces, then any of their closed subsets are also closed in the ambient space
and locally compact Hausdorff 
(cf.~\cite[Lemma~4.29]{lee_introduction_2000}).%
\end{remark}

In Section~\ref{sec:dynamic}, we develop a dynamic synergistic controller with the objective of globally asymptotically stabilizing a compact set for the {resulting} closed-loop system under the assumption that $\theta$ is known. In Section~\ref{sec:robust}, we modify the dynamic synergistic controller to allow for $\theta\in\Omega$ to be unknown, when $\Omega$ is known.

\section{{Dynamic Synergistic Hybrid Feedback}}\label{sec:dynamic}
\subsection{Controller Design}\label{sec:fundamentals}
Dynamic synergistic hybrid feedback (relative to the plant in Section~\ref{sec:setup}) is a hybrid control strategy that renders a compact set $\Amc\subset\Xmc\x\Xmc[c]$ globally asymptotically stable 
for the closed-loop system. It is comprised of a feedback law 
\begin{equation}
\label{eq:kappa}
\kappa:\dom\kappa\to\Umc
\end{equation}
and of the hybrid dynamics that are described in the sequel. 

Given a function
\begin{equation}
\label{eq:V}
\begin{aligned}
\V&:\dom\V\to\Rnneg\cup\{+\infty\},\\
\end{aligned}
\end{equation}
satisfying $\Xmc\x\Xmc[c]\subset\dom\V$ with $\dom\V$ open in the Euclidean space containing $\Xmc\x\Xmc[c]$,\footnote{The function $\V$ maps values in $\Xmc\x\Xmc[c]$ to the one-point compactification of $\Rnneg$. More generally, given a topological space $X$ that is noncompact locally compact Hausdorff space and an object $\infty$ not in $X$, the one point compactification of $X$ is a topological space $X^*$ with the topology:
$T=\{\text{open subsets of }X\}\cup
\{U\subset X^*: X^*\minus U \text{ is a compact subset of } X\}.$} and a set-valued map
\begin{equation}
\begin{aligned}
\label{eq:Q}
\jset[c]&:\Xmc\x\Xmc[c]\tto\Xmc[c]
\end{aligned}
\end{equation}
we define 
\begin{subequations}
\label{eq:mins}
\mathtoolsset{showonlyrefs=false}
\begin{align}
\label{eq:minV}\minV{\V}(x, \xc)&\ceq\min\{\V(x,g): g\in\jset[c](x,\xc)\},\\
\label{eq:qminV}\qminV{\V}(x, \xc)&\ceq\argmin\{\V(x,g):g\in\jset[c](x,\xc)\},\\
\label{eq:gap}\gap{\V}(x,\xc)&\ceq\V(x,\xc) -\minV{\V}(x, \xc)
\end{align}
\end{subequations}
for each $(x,\xc)\in\Xmc\x\Xmc[c]$, under the following assumption:
 \begin{enumerate}[label=(C\arabic*)]
\item\label{ass:Qbasic} The optimization problem in~\eqref{eq:mins} is feasible for each $(x,\xc)\in\Xmc\x\Xmc[c]$, i.e., for each $(x,\xc)\in\Xmc\x\Xmc[c]$, there exists $g\in\jset[c](x,\xc)$ such that $\V(x,g)<+\infty$.
 \setcounter{asscnt}{\value{enumi}}
 \end{enumerate}
 
Given a set-valued map $\F[c]$ {defined on} $\Xmc\x\Xmc[c]$ that {satisfies the following assumption:}
\begin{enumerate}[label=(C\arabic*)]
\setcounter{enumi}{\value{asscnt}}
\item \label{ass:Fc}{$F_c$} is outer semicontinuous, locally bounded, and convex-valued.
\setcounter{asscnt}{\value{enumi}}
\end{enumerate}
we define the hybrid controller dynamics as follows:%
\begin{subequations}
\label{eq:sysq}
\mathtoolsset{showonlyrefs=false}
\begin{align}
\label{eq:qflow}\dot{xc}& \in \F[c](x,\xc) & (x,\xc)&\in\fset\\
\label{eq:qjump} \xc\pl&\in\qminV{\V}(x,\xc) & (x,\xc)&\in\jset
\end{align}
\end{subequations}
where
\begin{equation}\label{eq:CD}
\begin{aligned}
\fset&\ceq\{(x,\xc)\in\Xmc\x\Xmc[c]:\gap{\V}(x,\xc)\leq\delta(x,\xc)\},\\
\jset&\ceq\{(x,\xc)\in\Xmc\x\Xmc[c]:\gap{\V}(x,\xc)\geq\delta(x,\xc)\},
\end{aligned}
\end{equation}
and $\delta:\Xmc\x\Xmc[c]\to\R{}$ is a continuous function.
The switching logic in~\eqref{eq:sysq} implements the following functionality: if the solutions to the closed-loop system reach a state $(x,\xc)$ where $\gap{\V}(x,\xc)$ is greater than or equal to the predefined value of $\delta(x,\xc)$, then the variable $\xc$ {is reset} to some point $g\in\qminV{\V}(x,\xc)$ and the feedback law changes its value from $\kappa(x,\xc)$ to $\kappa(x,g)$. Since the hybrid controller~\eqref{eq:sysq} is derived from $\kappa$, $\V$, $\jset[c]$, and $\fmap[c]$, we represent~\eqref{eq:sysq} using the $4$-tuple $(\kappa,\V,\jset[c],\fmap[c])$.

The hybrid closed-loop system $\Hmc\ceq(\fset,\fmap[cl],\jset,\jmap[cl])$ resulting from the interconnection between~\eqref{eq:sysx} and~$(\kappa,\V,\jset[c],\fmap[c])$ is given by
\hspace*{-6pt}\parbox{0.5\textwidth}{\begin{subequations}
\mathtoolsset{showonlyrefs=false}
\label{eq:closed}
\begin{align}
\pmtx{\dot x\\ \dot{xc}}&\in\fmap[cl](x,\xc)\ceq\pmtx{\F[\theta](x,\xc,\kappa(x,\xc))\\ \F[c](x,\xc)} & (x,\xc)\in\fset\\
\pmtx{x\pl \\ \xc\pl}&\in\jmap[cl](x,\xc)\ceq\pmtx{x\\ \qminV{\V}(x,\xc)} & (x,\xc)\in\jset.
\end{align}
\end{subequations}}

\begin{remark}\label{rem:delta}
    Notice that if $\delta(x,\xc)\geq 0$ for all $(x,\xc)\in\Xmc\x\Xmc[c]$ then it follows from the construction of the hybrid controller~$(\kappa,\V,\jset[c],\fmap[c])$ that $V(x,g)-V(x,\xc)\leq 0$ for each $(x,\xc)\in\jset$ and each $g\in\qminV{\V}(x,\xc)$. In other words, if the function $\delta$ is nonnegative for all $(x,\xc)\in\Xmc\x\Xmc[c]$, then the function $\V$ does not increase during jumps.
\end{remark}

The controller design presented in this section is informed by many preceding synergistic hybrid feedback controllers. As mentioned in Section~\ref{sec:contributions}, we preserve the switching logic of the synergistic controllers in~\cite[Chapter~7]{Sanfelice2021}, in the sense that controller switching is triggered when the difference between the current value of $\V$ and its lowest possible value exceeds a predefined threshold $\delta>0$. The main difference between the controller design presented in this paper and synergistic controllers in the literature is that, here, $\xc$ does not necessarily belong to a finite set. Instead, the flows of $\xc$ are described more generally by a differential inclusion and we constrain its jumps using a set-valued map $\jset[c]$.

In the sequel, we introduce the assumptions on the controller that allow for the global asymptotic stabilization of a compact subset of the state space.

\subsection{Basic Properties of the Closed-Loop System}\label{sec:basic}

In this section, we provide some conditions on~\eqref{eq:kappa},~\eqref{eq:V} and~\eqref{eq:Q} which ensure that the closed-loop system~\eqref{eq:closed} satisfies the hybrid basic conditions of~\cite[Assumption~6.5]{goebel_hybrid_2012} and that \hyperlink{def:maximal}{maximal} solutions to~\eqref{eq:closed} are complete.
To this end, we introduce the following definitions.

\begin{definition}\label{def:candidate}
Given a compact subset $\Amc$ of $\Xmc\x\Xmc[c]$, $\kappa$, $\V$, $\jset[c]$ and $\fmap[c]$ 
we say that the hybrid controller $(\kappa, \V, \jset[c],\fmap[c])$ is a synergistic candidate relative to $\Amc$ for~\eqref{eq:sysx} if {\ref{ass:Qbasic} and~\ref{ass:Fc} hold and}:
\begin{enumerate}[label=(C\arabic*)]
\setcounter{enumi}{\value{asscnt}}
\item\label{ass:V} $\V$ is continuous, positive definite relative to $\Amc$,\footnote{\hypertarget{def:pd}{A} function $\V:\Xmc\x\Xmc[c]\to\Rnneg$ is positive definite relative to $\Amc\subset\Xmc\x\Xmc[c]$ if $\V(x,\xc)=0\iff (x,\xc)\in\Amc$.} and $\V\inv([0,c])$ is compact for each $c\in\Rnneg$;%
\item\label{ass:Q} The set-valued map $\jset[c]$ is outer semicontinuous, lower semicontinuous, and locally bounded;%
\item\label{ass:kappa} The function $\kappa$ is continuous and $$\{(x,\xc)\in\Xmc\x\Xmc[c]:\V(x,\xc)<+\infty\}\subset\dom\kappa.$$
\setcounter{asscnt}{\value{enumi}}
\end{enumerate}
\end{definition}

Given a synergistic candidate relative to $\Amc$, the property~\ref{ass:V} guarantees that sublevel sets of $\V$ are compact and the properties~\ref{ass:V} and~\ref{ass:Q} guarantee that the synergy gap function $\gap{\V}$ in~\eqref{eq:gap} is continuous and that $\qminV{\V}$ is outer semicontinuous, as proved in the next result.

\begin{lemma}\label{lem:regular}
Given a compact subset $\Amc$ of $\Xmc\x\Xmc[c]$, if~$(\kappa,\V,\jset[c],\fmap[c])$ is a synergistic candidate relative to $\Amc$ for~\eqref{eq:sysx}, then the following hold:
\begin{enumerate}
\item The function $\minV{\V}$ in~\eqref{eq:minV} is continuous;
\item The set-valued map $\qminV{\V}$ in~\eqref{eq:qminV} is outer semicontinuous and $\qminV{\V}(x,\xc)$ is compact for each $(x,\xc)\in\Xmc\x\Xmc[c]$;
\item The function $\gap{\V}$ in~\eqref{eq:gap} is continuous.
\end{enumerate}
\end{lemma}
\begin{proof}
It follows from~\ref{ass:Q} that $\jset[c]$ is outer semicontinuous, hence $\jset[c](x,\xc)$ is closed for each $(x,\xc)\in\Xmc\x\Xmc[c]$. Since $\jset[c]$ is also assumed to be locally bounded in~\ref{ass:Q}, we have that $\jset[c](x,\xc)$ is compact for each $(x,\xc)\in\Xmc\x\Xmc[c]$. In addition, the outer semicontinuity and local boundedness of $\jset[c]$ imply that $\jset[c]$ is upper semicontinuous (cf.~\cite[Lemma~5.15]{goebel_hybrid_2012}). Since $\jset[c]$ is assumed to be lower semicontinuous in~\ref{ass:Q}, we have that $\jset[c]$ is continuous.
Since $\V$ is continuous by assumption~\ref{ass:V}, it follows from~\cite[Theorem~9.14]{Sundaram1996} that $\minV{\V}$ is continuous and that $\qminV{\V}$ is compact-valued and upper semicontinuous. Since $\Xmc$ is locally compact Hausdorff (cf. Remark~\ref{rem:hausdorff}), it follows from~\cite[Proposition~4.27]{lee_introduction_2000} that each point $(x,\xc)\in\Xmc\x\Xmc[c]$ has a precompact neighborhood $U_x$. Since $\qminV{\V}$ is compact-valued and upper semicontinuous, it follows from~\cite[Proposition~9.7]{Sundaram1996} that $\qminV{\V}(\cl{U_x})$ is compact. It follows from the fact that $\qminV{\V}(U_x)$ is a subset of a the compact set $\qminV{\V}(\cl{U_x})$ that $\qminV{\V}$ is locally bounded. Since $\qminV{\V}$ is compact-valued it is, in particular, closed-valued, hence it follows from~\cite[Lemma~5.15]{goebel_hybrid_2012} that $\qminV{\V}$ is outer semicontinuous. 
It follows from~\ref{ass:Qbasic} that $\minV{\V}(x,\xc)<+\infty$ for each $(x,\xc)\in\Xmc\x\Xmc[c]$, hence the function $\gap{\V}$ is continuous because it is the composition of continuous functions. 
\end{proof}

The hybrid basic conditions in~\cite[Assumption~6.5]{goebel_hybrid_2012} are very important to the synthesis of hybrid controllers, because they guarantee that the resulting hybrid closed-loop systems are endowed with nominal robustness to a wide range of perturbations/sensor noise and, in particular, they enable the application of invariance principles for hybrid systems (cf.~\cite[Chapter~8]{goebel_hybrid_2012}). In the following result, we show that these conditions follow directly from the regularity of~\eqref{eq:gap} and~\eqref{eq:mins} that was proved in Lemma~\ref{lem:regular}.

\begin{corollary}\label{cor:hbc}
{Suppose that Assumption~\ref{ass:data} holds. }Given a compact subset $\Amc$ of $\Xmc\x\Xmc[c]$, if~$(\kappa,\V,\jset[c],\fmap[c])$ is a synergistic candidate relative to $\Amc$ for~\eqref{eq:sysx}, then the hybrid closed-loop system~\eqref{eq:closed} satisfies~\ref{ass:basicCD},~\ref{ass:basicF}, and~\ref{ass:basicG}.
\end{corollary}
\begin{proof}
Let $h(x,\xc)\ceq \gap{\V}(x,\xc)-\delta(x,\xc)$ for each $(x,\xc)\in\Xmc\x\Xmc[c]$. Since $\delta$ is assumed to be continuous and $\gap{\V}$ is continuous under the given assumptions (cf.~Lemma~\ref{lem:regular}), it follows that $h$ is a continuous function. Continuity of $h$ implies that the flow and jump sets are closed, because they can be written as the preimage of closed sets through $h$, as follows: $\fset=h\inv((-\infty,0])$ and $\jset=h\inv([0,+\infty])$, respectively (cf.~\cite[Lemma~2.7]{lee_introduction_2000}).

It follows from the construction of the hybrid controller~$(\kappa,\V,\jset[c],\fmap[c])$ that $\fmap[cl](x,\xc)$ is defined for each $(x,\xc)\in\fset$. From the continuity of $\kappa$ in~\ref{ass:kappa} and the assumption that $\F[\theta]$ is outer semicontinuous, locally bounded and convex-valued (cf.~\ref{ass:Fbasic}), it follows that $\fmap[cl]$ in~\eqref{eq:closed} is outer semicontinuous and locally bounded relative to $\fset$ and $\fmap[cl](x,\xc)$ is convex for each $(x,\xc)\in\fset$.
The outer semicontinuity and local boundedness of $\jmap[cl]$ in~\eqref{eq:closed} relative to $\jset$ follows from Lemma~\ref{lem:regular}.
\end{proof}


\subsection{Global Asymptotic Stability of $\Amc$}\label{sec:gas}

In this section, we present further assumptions on the hybrid controller~$(\kappa,\V,\jset[c],\fmap[c])$ that allow for the global asymptotic stabilization of a compact set $\Amc$ for~\eqref{eq:closed}.

\begin{definition}
Given a compact subset $\Amc$ of $\Xmc\x\Xmc[c]$, we say that a synergistic candidate relative to $\Amc$ for~\eqref{eq:sysx} with data $(\kappa,\V,\jset[c],\fmap[c])$, is synergistic relative to $\Amc$ for~\eqref{eq:sysx} if:
\begin{enumerate}[label=(C\arabic*)]
\setcounter{enumi}{\value{asscnt}}
\item\label{ass:uc} The function $\V$ is Lipschitz continuous on a neighborhood of $C$ and the growth of $\V$ along flows of~\eqref{eq:closed} is bounded by $\uc$ with%
\begin{align*}
\uc(x,\xc)&\leq 0  & &\forall (x,\xc)\in\Xmc\x\Xmc[c];
\end{align*}
\item\label{ass:Psi} The largest weakly invariant subset of
\begin{equation}
\label{eq:flows}
\begin{aligned}
\dot x&\in {\fmap[\theta]}(x,\xc,\kappa(x,\xc))\\
\dot{xc}&\in \F[c](x,\xc)
\end{aligned}
\end{equation}
in $\cl{\uc\inv(0)}$, denoted by $\Psi$, is such that 
\begin{equation}\label{eq:deltabnd}
\deltabnd[1]\ceq\inf\{\gap{\V}(x,\xc):(x,\xc)\in\Psi\minus\Amc\}>0.
\end{equation}
\setcounter{asscnt}{\value{enumi}}
\end{enumerate}
\end{definition}

If one considers $\V$ as a Lyapunov function candidate, then Assumption~\ref{ass:uc} implies that $\V$ is nonincreasing along flows to the closed-loop system~\eqref{eq:closed}, implying that there exists a choice of $\delta$ which renders $\Amc$ stable for~\eqref{eq:closed}. 

\begin{lemma}\label{lem:complete}
    {Suppose that Assumption~\ref{ass:data} holds. }Given a compact subset $\Amc$ of $\Xmc\x\Xmc[c]$, if~$(\kappa,\V,\jset[c],\fmap[c])$ is a synergistic candidate relative to $\Amc$ for~\eqref{eq:sysx} satisfying~\ref{ass:uc} and $\delta(x,\xc)\geq 0$ for each $(x,\xc)\in\Xmc\x\Xmc[c]$, then each sublevel set of $\V$ is \hyperlink{def:invariant}{forward pre-invariant} for~\eqref{eq:closed}. If, for each $(x,\xc)\in\fset\minus\jset$,
    \begin{enumerate}[label=(VC)]
    \item\label{ass:VC} there exists a neighborhood $U$ of $(x,\xc)$ such that 
    $\fmap[cl](\xi)\cap \T[\xi]\fset\neq\emptyset,$
    for every $\xi\in U\cap\fset$
    \end{enumerate}
    then each maximal solution to~\eqref{eq:closed} is complete and, consequently, each sublevel set of $\V$ is forward invariant.
\end{lemma}
\begin{proof}
It follows from the discussion in Remark~\ref{rem:delta} that the growth of $\V$ along jumps of~\eqref{eq:closed} is bounded by $\ud$ with
\begin{equation}
\label{eq:ud}
\ud(x,\xc)\leq\begin{cases}
-\delta(x,\xc) & \text{ if } (x,\xc)\in\jset\\
-\infty & \text{otherwise}
\end{cases}
\end{equation}
for each $(x,\xc)\in\Xmc\x\Xmc[c]$.
Together with assumption~\ref{ass:uc} it follows that the growth of $\V$ along solutions to~\eqref{eq:closed} is bounded by $\uc,\ud$ satisfying
\begin{equation}
    \label{eq:ucud}
\begin{aligned}
\uc(x,\xc)&\leq 0, & \ud(x,\xc)&\leq 0
\end{aligned}
\end{equation}
for each $(x,\xc)\in\Xmc\x\Xmc[c]$. It follows that each solution $\phi$ to~\eqref{eq:closed} with initial condition $\xi$ satisfies $\V(\phi(t,j))\leq\V(\xi)$ for all $(t,j)\in\dom\phi$, hence each sublevel set of $\V$ is forward pre-invariant for~\eqref{eq:closed}.

It follows from Corollary~\ref{cor:hbc} that~\eqref{eq:closed} satisfies the hybrid basic conditions, hence we can use~\cite[Proposition~6.10]{goebel_hybrid_2012} to prove the completeness of each maximal solution to~\eqref{eq:closed}.
Since $\fset\cup\jset=\Xmc\x\Xmc[c]$, then there are no solutions to~\eqref{eq:closed} starting outside the union between the jump and flow sets. It follows from~\ref{ass:VC} that (VC) in~\cite[Proposition~6.10]{goebel_hybrid_2012} is satisfied, hence each maximal solution to~\eqref{eq:closed} either ``blows up,'' leaves $\fset\cup\jset$ in finite time or is complete (cf. conditions \textit{(a)},\textit{(b)} and~\textit{(c)} of~\cite[Proposition~6.10]{goebel_hybrid_2012}).
Since $\jmap[cl](\jset)\subset\fset\cup\jset$, no solution can leave $\fset\cup\jset$ {after a jump} (hence, condition (c) in~\cite[Proposition~6.10]{goebel_hybrid_2012} does not occur). Since each sublevel set of $\V$ is compact and forward pre-invariant, then solutions to~\eqref{eq:closed} do not ``blow up'' (condition{ (b) in}~\cite[Proposition~6.10]{goebel_hybrid_2012} does not occur). It follows that each maximal solution to~\eqref{eq:closed} is complete.
\end{proof}

\begin{lemma}\label{lem:stability}
    {Suppose that Assumption~\ref{ass:data} holds. }Given a compact subset $\Amc$ of $\Xmc\x\Xmc[c]$, if~$(\kappa,\V,\jset[c],\fmap[c])$ is a synergistic candidate relative to $\Amc$ for~\eqref{eq:sysx} that satisfies~\ref{ass:uc} and $\delta(x,\xc)\geq 0$ for each $(x,\xc)\in\Xmc\x\Xmc[c]$, then the set $\Amc$ is stable for~\eqref{eq:closed}.%
\end{lemma}
\begin{proof}
Since $\Xmc\x\Xmc[c]\subset\dom\V$, it follows that $\gap{\V}(x,\xc)$ is defined for all $(x,\xc)\in\Xmc\x\Xmc[c]$; hence, for any given continuous function $\delta:\Xmc\x\Xmc[c]\to\R{}$, at least one of the following conditions holds:
\begin{inparaenum}
    \item $\gap{\V}(x,\xc)\geq\delta(x,\xc)$;
    \item $\gap{\V}(x,\xc)\leq\delta(x,\xc)$.
\end{inparaenum}
It follows from~\eqref{eq:CD} that $\cl{\fset}\cup\jset=\Xmc\x\Xmc[c]\subset\dom\V$. Since $\dom\V$ is also assumed to be open in the Euclidean space containing $\Xmc\x\Xmc[c]$, it follows that $\dom\V$ contains a neighborhood of $\Amc\cap\left(\fset\cup\jset\cup\jmap(\jset)\right)$.
Positive definiteness of $\V$ with respect to $\Amc$ and continuity of $\V$ follows from~\ref{ass:V}. From assumption~\ref{ass:uc} and from~\eqref{eq:ud}, it follows that $\V$ is locally Lipschitz on a neighborhood of $\cl{\fset}$ and that the bounds~\cite[Eqs.(3.18),~(3.19)]{Sanfelice2021} are satisfied. Since $\Amc$ is compact and the hybrid basic conditions are satisfied (cf. Corollary~\ref{cor:hbc}), it follows from~{\cite[Theorem~3.19]{Sanfelice2021}} that $\Amc$ is stable for~\eqref{eq:sysx}.
\end{proof}

Assumption~\ref{ass:Psi} guarantees that there exists $\delta$ satisfying
\begin{enumerate}[label=(D\arabic*)]
    \item\label{ass:deltazero} $\delta(x,\xc)>0$ for each $(x,\xc)\in\Xmc\x\Xmc[c]$;
    \item\label{ass:deltapsi} $\delta(x,\xc)<\gap{\V}(x,\xc)$ for each $(x,\xc)\in\Psi\minus\Amc$ with $\Psi$ defined in~\ref{ass:Psi}.
    \setcounter{deltacnt}{\value{enumi}}
    \end{enumerate}
\noindent We say that $\delta$ is positive if it satisfies~\ref{ass:deltazero}, and that a hybrid controller~$(\kappa,\V,\jset[c],\fmap[c])$ is synergistic relative to $\Amc$ for~\eqref{eq:sysx} with synergy gap exceeding $\delta$ if it is synergistic relative to $\Amc$ for~\eqref{eq:sysx} and satisfies~\ref{ass:deltapsi}. When both conditions~\ref{ass:deltazero} and~\ref{ass:deltapsi} are satisfied, all the points in the largest weakly invariant subset of~\eqref{eq:flows} in $\cl{\uc\inv(0)}$ that are not in $\Amc$ lie in the jump set of~\eqref{eq:closed}, allowing us to prove that $\Amc$ is globally asymptotically stable for the closed-loop system~\eqref{eq:closed}.

\begin{theorem}\label{thm:gas}
    {Suppose that Assumption~\ref{ass:data} holds. }Given a compact subset $\Amc$ of $\Xmc\x\Xmc[c]$ and a positive function $\delta:\Xmc\x\Xmc[c]\to\R{}$, if~$(\kappa,\V,\jset[c],\fmap[c])$ is synergistic relative to $\Amc$ for~\eqref{eq:sysx} with synergy gap exceeding $\delta$,
then the set $\Amc$ is globally pre-asymptotically stable for~\eqref{eq:closed}. If, for each $(x,\xc)\in\fset\minus\jset$,~\ref{ass:VC} is satisfied, then $\Amc$ is globally asymptotically stable for~\eqref{eq:closed}.
\end{theorem}
\begin{proof}
Stability of $\Amc$ is proved in Lemma~\ref{lem:stability} and completeness of solutions is demonstrated in Lemma~\ref{lem:complete}. The global pre-asymptotic stability of $\Amc$ for~\eqref{eq:closed} follows from pre-attractivity of $\Amc$ for~\eqref{eq:closed}, which is demonstrated next through an application of~\cite[Theorem~8.2]{goebel_hybrid_2012}.

It follows from Lemmas~\ref{lem:stability} and~\ref{lem:complete} that each maximal solution to~\eqref{eq:closed} is precompact and the growth of $\V$ along solutions to~\eqref{eq:closed} is bounded by $\uc,\ud$ satisfying~\eqref{eq:ucud}. Therefore, it follows from~\cite[Theorem~8.2]{goebel_hybrid_2012} that every complete solution approaches the largest weakly invariant set 
\begin{equation}
\label{eq:8_2_set}
\V\inv(r)\cap\left(\cl{\uc\inv(0)}\cup(\ud\inv(0)\cap\jmap[cl](\ud\inv(0))\right)
\end{equation}
for some $r$ in the image of $\V$. From~\eqref{eq:ud} and the assumption~\ref{ass:deltazero}, it follows that $\ud\inv(0)=\emptyset$, hence~\eqref{eq:8_2_set} can be rewritten as
\begin{equation}
\label{eq:8_2_set2}
\V\inv(r)\cap\cl{\uc\inv(0)}.
\end{equation}
It follows from~\ref{ass:Psi},~\ref{ass:deltapsi} and the definition of $\jset$ in~\eqref{eq:sysq} that the largest weakly invariant subset of~\eqref{eq:closed} in~\eqref{eq:8_2_set2} does not include points that are not in $\Amc$ and, consequently, $\Amc$ is globally pre-attractive for~\eqref{eq:closed}. Global asymptotic stability of $\Amc$ for~\eqref{eq:closed} follows from global pre-asymptotic stability if each maximal solution to~\eqref{eq:closed} is complete, which is guaranteed by Lemma~\ref{lem:complete} under assumption~\ref{ass:VC}.
\end{proof}

Note that, if there exists an accumulation point of $\Psi\minus\Amc$ in $\Amc$, then $\deltabnd[1]$ in~\eqref{eq:deltabnd} is equal to $0$. Therefore, the topology of $\Psi$ and $\Amc$ may preclude global asymptotic stabilization of $\Amc$ for~\eqref{eq:closed} since~\ref{ass:Psi} is not met. Conversely, if one is able to show that $\deltabnd[1]>0$, then $\Psi\minus\Amc$ does not have accumulation points in $\Amc$. With additional conditions on $\delta$, we are able to show that there exists a neighborhood of $\Amc$ contained in $\fset$.

\begin{proposition}\label{pro:AmcLow}
Given a compact subset $\Amc$ of $\Xmc\x\Xmc[c]$, if the hybrid controller~$(\kappa,\V,\jset[c],\fmap[c])$ is synergistic relative to $\Amc$ and
$\deltalow\ceq\inf\{\delta(x,\xc):(x,\xc)\in\Xmc\x\Xmc[c]\}$
satisfies $\deltalow\in(0,\deltabnd[1])$, where $\deltabnd[1]$ is given in~\ref{ass:Psi}, then there exists a neighborhood of $\Amc$ that is contained in $\fset$.
\end{proposition}
\begin{proof}
Selecting $\epsilon\in(0,\deltalow)$, we have that $\gap{\V}\inv((-\epsilon,\epsilon))$ is open because $\gap{\V}$ is continuous (cf. Lemma~\ref{lem:regular}), contains $\Amc$ because $\gap{\V}(\Amc)=0$ and it is a subset of $\fset$ because $\epsilon<\deltalow\leq\delta(x,\xc)$ for each $(x,\xc)\in\Xmc\x\Xmc[c]$.
\end{proof}

\begin{remark}
Note that, if the hybrid controller~$(\kappa,\V,\jset[c],\fmap[c])$ is synergistic relative to $\Amc$ for~\eqref{eq:sysx}, then $\delta$ can be chosen as a constant $\Delta\in\R{}$ as long as $\Delta\in (0,\deltabnd[1])$.
 In this case, the conditions of Proposition~\ref{pro:AmcLow} hold, thus the fact that $\delta$ is state-dependent does not constrain the global asymptotic stability results and it provides more flexibility to the design of the hybrid controller.
\end{remark}

\section{Robust Synergistic Hybrid Feedback}\label{sec:robust}
In this section, we propose a new kind of synergistic hybrid controller that, unlike the controller of Section~\ref{sec:dynamic}, is able to handle the case where $\theta$ is unknown, but belongs to a known compact set $\Omega$.
In this direction, let $\Ascr\ceq\{\Amc[\theta]\}_{\theta\in\Omega}$ denote a collection of compact subsets of $\Xmc\x\Xmc[c]$ and let $\Vscr\ceq\{\V[\theta]\}_{\theta\in\Omega}$ denote a collection of functions satisfying the following assumption:
\begin{enumerate}[label=(C\arabic*)]
    \setcounter{enumi}{\value{asscnt}}
    \item\label{ass:Vtheta} Given a compact set $\Omega$ and a collection $\Vscr\ceq\{\V[\theta]\}_{\theta\in\Omega}$ of functions $\V[\theta]:\dom\V[\theta]\to\Rnneg\cup\{+\infty\}$ satisfying $\Xmc\x\Xmc[c]\subset\dom\V[\theta]$ for each $\theta\in\Omega$, we assume that
    \begin{equation}
        \label{eq:VtoVtheta}
        (x,\xc,\theta)\mapsto\V(x,\xc,\theta)\ceq\V[\theta](x,\xc)
    \end{equation}
    is continuous.
    \setcounter{asscnt}{\value{enumi}}
 \end{enumerate}
 \begin{remark}
Note that $\Omega$ might be uncountable, thus the collections $\Ascr\ceq\{\Amc[\theta]\}_{\theta\in\Omega}$ and $\Vscr\ceq\{\V[\theta]\}_{\theta\in\Omega}$ are not necessarily finite nor countable.
\end{remark}

Using the previous definitions, we propose the following hybrid controller: 
\begin{subequations}
\label{eq:rsysq}
\mathtoolsset{showonlyrefs=false}
\begin{align}
\label{eq:rqflow}\dot{xc} &\in \F[c](x,\xc) &(x,\xc)&\in\fset[\Omega]\\
\label{eq:rqjump} 
\xc\pl&\in\jmap[c](x,\xc) &  (x,\xc)&\in\jset[\Omega]
\end{align}
\end{subequations}
where 
\begin{equation}
    \label{eq:D_Omega}
\begin{aligned}
\fset[\Omega]&\ceq\left\{(x,\xc)\in\Xmc\x\Xmc[c]:\min_{\theta\in\Omega}\gap{\V[\theta]}(x,\xc)\leq\delta(x,\xc)\right\}\\
\jset[\Omega]&\ceq\left\{(x,\xc)\in\Xmc\x\Xmc[c]:\min_{\theta\in\Omega}\gap{\V[\theta]}(x,\xc)\geq\delta(x,\xc)\right\}
\end{aligned}
\end{equation}
with $\delta:\Xmc\x\Xmc[c]\to\R{}$ continuous, and%
\begin{equation*}
\begin{multlined}
\min_{\theta\in\Omega}\gap{\V[\theta]}(x,\xc) = \min_{\theta\in\Omega} \left\{\V[\theta](x,\xc) -\minV{{\V[\theta]}}(x, \xc)\right\}\\
= \min_{\theta\in\Omega} \left\{\V(x,\xc,\theta) -\min_{g\in\jset[c](x,\xc)}\V(x,g,\theta)\right\}
\end{multlined}
\end{equation*}%
for each $(x,\xc)\in\Xmc\x\Xmc[c]$, in accordance with the definitions~\eqref{eq:gap},~\eqref{eq:minV} and~\eqref{eq:VtoVtheta}, and $\jmap[c]:\Xmc\x\Xmc[c]\tto\Xmc$  satisfies the following assumptions:
\begin{enumerate}[label=(C\arabic*)]
    \setcounter{enumi}{\value{asscnt}}
    \item\label{ass:Gc_regular} The set-valued map $\jmap[c]$ is outer semicontinuous and locally bounded;
    \item\label{ass:Gc_decreases} For each $\theta\in\Omega$, we assume that
    \begin{equation}\label{eq:Gtt} 
        \V[\theta](x,\xc)-\V[\theta](x,g)\geq \min_{\theta\in\Omega}\gap{\V[\theta]}(x,\xc)
    \end{equation}
    for each $(x,\xc)\in\jset[\Omega]$ and each $g\in\jmap[c](x,\xc)$
    \setcounter{asscnt}{\value{enumi}}
\end{enumerate} 
\begin{remark}\label{rem:Gc}
Under assumption~\ref{ass:Gc_decreases}, we guarantee by construction that $$\V[\theta](x,\xc)-\V[\theta](x,g)\geq \delta(x,\xc)$$ for each $(x,\xc)\in\jset[\Omega]$ and each $g\in\jmap[c](x,\xc)$, which implies that the function $\V[\theta]$ does not increase during jumps if $\delta(x,\xc)\geq 0$ for all $(x,\xc)\in\Xmc\x\Xmc[c]$ (cf.~Remark~\ref{rem:delta}). 
\end{remark}
Note that the jump map of the hybrid controller~\eqref{eq:sysq} is constructed from the data $\V$ and $\jset[c]$ as shown in~\eqref{eq:qminV}. On the other hand, the jump map $\jmap[c]$ in~\eqref{eq:rsysq} is left undefined for the sake of generality.
It is possible to construct $\jmap[c]$ in~\eqref{eq:rsysq} from the data $\Vscr$ and $\jset[c]$, but this requires additional assumptions, as shown in the following remark. Owing to the fact that~\eqref{eq:rsysq} is derived from $\kappa$, $\Vscr$, $\jset[c]$, $\fmap[c]$, and $\jmap[c]$, we refer to~\eqref{eq:rsysq} using the $5$-tuple $(\kappa, \Vscr, \jset[c],\fmap[c],\jmap[c])$.

\begin{remark}\label{rem:Gtt}
Suppose that $\Omega$ is compact and convex and that $\jset[c]$ in~\eqref{eq:Q} is convex and compact for each $(x,\xc)\in\Xmc\x\Xmc[c]$. Given $\Vscr\ceq\{\V[\theta]\}_{\theta\in\Omega}$, suppose that $\jmap[c]$ defined as%
\begin{multline}\label{eq:maxmin}
\jmap[c](x,\xc)\ceq\argmax_{g\in\jset[c](x,\xc)}\quad \min_{\theta\in\Omega}\left\{ \V(x,\xc,\theta)-\V(x,g,\theta)\right\}\\ \forall(x,\xc)\in\Xmc\x\Xmc[c]
\end{multline}
is outer semicontinuous with $\V(x,\xc,\theta)=\V[\theta](x,\xc)$ for each $(x,\xc,\theta)\in\Xmc\x\Xmc[c]\x\Omega$. Furthermore, suppose that, for each $(x,\xc)\in\Xmc\x\Xmc[c]$, the function $h(g,\theta)\ceq \V[\theta](x,\xc)-\V[\theta](x,g)$
is continuous, quasi-concave as a function of $g$, and quasi-convex as a function of $\theta$.%
\footnote{A function $(x,y)\mapsto h(x,y)$ on $X\x Y$ is quasi-concave as a function of $x$ if the set $\{x\in X:h(x,y)\geq c\}$ is convex for each $y\in Y$ and each $c\in\R{}$. The function $h$ is quasi-convex as a function of $y$ if the set $\{y\in X:h(x,y)\leq c\}$ is convex for each $x\in X$ and each $c\in\R{}$.}
Then, it follows from~\cite[Theorem~3.4]{sion_general_1958} that the $\min$ and $\max$ operations in $\max_{g\in\jset[c](x,\xc)}\min_{\theta\in\Omega}h(g,\theta)$ commute, yielding
\begin{equation}\label{eq:haux}
\begin{multlined}
\max_{g\in\jset[c](x,\xc)}\min_{\theta\in\Omega}h(g,\theta)=
    \min_{\theta\in\Omega}\max_{g\in\jset[c](x,\xc)}h(g,\theta) \\
    =\min_{\theta\in\Omega} \left\{\V[\theta](x,\xc)-\min_{g\in\jset[c](x,\xc)}\V[\theta](x,g)\right\}.
\end{multlined}
\end{equation}
It follows from~\eqref{eq:haux} and~\eqref{eq:gap} that
$
\max_{g\in\jset[c](x,\xc)}\min_{\theta\in\Omega}h(g,\theta)
    =\min_{\theta\in\Omega}\gap{\V[\theta]}(x,\xc).
$
We conclude that, for each $\theta\in\Omega$, the following holds
$\V[\theta](x,\xc)-\V[\theta](x,g)\geq\min_{\theta\in\Omega}\gap{\V[\theta]}(x,\xc),$
for each $(x,\xc)\in\Xmc\x\Xmc[c]$ and each $g$ belonging to~\eqref{eq:maxmin}, hence condition~\eqref{eq:Gtt} is verified.
\end{remark}

The following definition extends the notion of a synergistic controller in order to address the case where $\theta\in\Omega$ is not known. 

\begin{definition}\label{def:robust}
Given a compact set $\Omega$, a continuous function $\delta:\Xmc\x\Xmc[c]\to\R{}$, a collection of compact subsets $\Ascr\ceq\{\Amc[\theta]\}_{\theta\in\Omega}$ of $\Xmc\x\Xmc[c]$, and a collection of continuous functions $\Vscr\ceq\{\V[\theta]\}_{\theta\in\Omega}$, we say that the hybrid controller~$(\kappa,\Vscr,\jset[c],\fmap[c],\jmap[c])$ is \emph{synergistic relative to $\Ascr$ for~\eqref{eq:sysx} with robustness margin $\Omega$} if~\ref{ass:Vtheta},~\ref{ass:Gc_regular},~\ref{ass:Gc_decreases} hold, and if, for each $\theta\in\Omega$, the hybrid controller~$(\kappa,\V[\theta],\jset[c],\fmap[c])$ is synergistic relative to $\Amc[\theta]$ for~\eqref{eq:sysx}.
\end{definition}

The assumption that the hybrid controller~$(\kappa,\Vscr,\jset[c],\fmap[c],\jmap[c])$ is synergistic relative to $\Ascr$ for~\eqref{eq:sysx} with robustness margin $\Omega$ ensures that the hybrid closed-loop system 
\begin{subequations}
\mathtoolsset{showonlyrefs=false}
\label{eq:rclosed}
\begin{align}
\pmtx{\dot x\\ \dot{xc}}&\in\fmap[cl](x,\xc)\ceq\pmtx{\F[\theta](x,\xc,\kappa(x,\xc))\\ \F[c](x,\xc)} & (x,\xc)\in\fset[\Omega]\\
\pmtx{x\pl \\ \xc\pl}&\in\jmap[\Omega](x,\xc)\ceq\pmtx{x\\ \jmap[c](x,\xc)} & (x,\xc)\in\jset[\Omega]
\end{align}
\end{subequations}
satisfies the hybrid basic conditions as proved next. 

\begin{lemma}\label{lem:rhbc}
Suppose that Assumption~\ref{ass:data} holds. Given a compact set $\Omega$ and a collection of compact subsets $\Ascr\ceq\{\Amc[\theta]\}_{\theta\in\Omega}$ of $\Xmc\x\Xmc[c]$ if~$(\kappa,\Vscr,\jset[c],\fmap[c],\jmap[c])$ is synergistic relative to $\Ascr$ for~\eqref{eq:sysx} with robustness margin $\Omega$, then the hybrid closed-loop system~\eqref{eq:rclosed} satisfies~\ref{ass:basicCD},~\ref{ass:basicF}, and \ref{ass:basicG}.
\end{lemma}
\begin{proof}
The continuity of $\gap{\V[\theta]}$ (for a fixed $\theta\in\Omega$) is established in Lemma~\ref{lem:regular}.
It follows from the continuity of $(x,\xc,\theta)\mapsto \V(x,\xc,\theta)=\V[\theta](x,\xc)$ that is assumed in~\ref{ass:Vtheta}, compactness of $\Omega$ and from~\cite[Theorem~9.14]{Sundaram1996} that the function 
\begin{equation}\label{eq:gaptt}
(x,\xc)\mapsto\min_{\theta\in\Omega}\gap{\V[\theta]}(x,\xc)
\end{equation}
is continuous on $\Xmc\x\Xmc[c]$.
It follows from the continuity of~\eqref{eq:gaptt} and of $\delta$ that $\fset[\Omega]$ and $\jset[\Omega]$ are closed, because they are the preimage of the closed sets $(-\infty,0]$ and $[0,+\infty]$, respectively.
It follows from Assumption~\ref{ass:data},~\ref{ass:Fc}\add{,} and~\ref{ass:kappa} that the flow map $\F[\Omega]$ is outer semicontinuous, locally bounded and convex-valued.
It follows from \ref{ass:Gc_regular} that $\jmap[\Omega]$ is outer semicontinuous\add{,} and locally bounded relative to $\jset[\Omega]$.
\end{proof}

In the sequel, we demonstrate that, for each $\theta\in\Omega$, the set $\Amc[\theta]\in\Ascr$ is globally asymptotically stable under appropriate assumptions on $\delta$. The next result asserts forward pre-invariance of sublevel sets of $\V[\theta]\in\Vscr$ for the closed-loop system~\eqref{eq:rclosed} when $\delta$ is a continuous and nonnegative function. 

\begin{lemma}\label{lem:rstability}
Suppose that Assumption~\ref{ass:data} holds. Given a compact set $\Omega$ and a collection of compact subsets $\Ascr\ceq\{\Amc[\theta]\}_{\theta\in\Omega}$ of $\Xmc\x\Xmc[c]$, if~$(\kappa,\Vscr,\jset[c],\fmap[c],\jmap[c])$ is synergistic relative to $\Ascr$ for~\eqref{eq:sysx} with robustness margin $\Omega$ and if $\delta(x,\xc)\geq 0$ for each $(x,\xc)\in\Xmc\x\Xmc[c]$, then, for each $\theta\in\Omega$, each sublevel set of $\V[\theta]$ is forward pre-invariant for~\eqref{eq:rclosed}.
If, for each $(x,\xc)\in\fset[\Omega]\minus\jset[\Omega]$,
\begin{enumerate}[label=(VC')]
\item\label{ass:VC2} there exists a neighborhood $U$ of $(x,\xc)$ such that 
$\fmap[cl](\xi)\cap \T[\xi]\fset[\Omega]\neq\emptyset,$
for every $\xi\in U\cap\fset[\Omega]$
\end{enumerate}
then each maximal solution to~\eqref{eq:rclosed} is complete and, consequently, each sublevel set of $\V[\theta]$ is forward invariant for~\eqref{eq:rclosed}. 
\end{lemma}
\begin{proof}
As explained in Remark~\ref{rem:Gc}, it follows from~\ref{ass:Gc_decreases} and~\eqref{eq:D_Omega} that 
$\V[\theta](x,\xc)-\V[\theta](x,g)\geq\delta(x,\xc)$
for each $g\in\jmap[c](x,\xc)$ and each $(x,\xc)\in\jset[\Omega]$.
Hence, the growth of $\V[\theta]$ during jumps of~\eqref{eq:rclosed} is bounded by
\begin{equation}
\label{eq:udtt}
\ud[theta](x,\xc)\ceq\begin{cases}
-\delta(x,\xc) &\text{ if }(x,\xc)\in\jset[\Omega]\\
-\infty & \text{ otherwise }
\end{cases}
\end{equation}
for each $(x,\xc)\in\Xmc\x\Xmc[c]$. Since~$(\kappa,\Vscr,\jset[c],\fmap[c],\jmap[c])$ is synergistic relative to $\Ascr$ for~\eqref{eq:sysx} with robustness margin $\Omega$, it follows that~$(\kappa,\V[\theta],\jset[c],\fmap[c])$ is synergistic relative to $\Amc[\theta]$ for~\eqref{eq:sysx} and, due to this assumption, the remainder of the proof follows closely that of Lemma~\ref{lem:complete}. 
From Assumption~\ref{ass:uc} it follows that the growth of $\V[\theta]$ along solutions to~\eqref{eq:rclosed} is bounded by $\uc[theta],\ud[theta]$, with $\uc[theta](x,\xc)\leq 0$ and $\ud[theta](x,\xc)\leq 0$
for each $(x,\xc)\in\Xmc\x\Xmc[c]$. This implies that sublevel sets of $\V[\theta]$ are forward pre-invariant for~\eqref{eq:rclosed}. The completeness of solutions under~\ref{ass:VC2} follows closely the proof in Lemma~\ref{lem:complete}, thus it is omitted here.
\end{proof}

Let $\Psi[\theta]$ denote the largest weakly invariant subset of
\begin{align*}
    (\dot{x},\dot{xc})&\in \fmap[cl](x,\xc) & & (x,\xc)\in\bar{\uc[theta]\inv(0)}
\end{align*}
where $\uc[theta]$ is the upper bound on the growth of $\V[\theta]$ during flows of~\eqref{eq:rclosed} as defined in~\ref{ass:uc}. Given a function $\delta:\Xmc\x\Xmc[c]\to\R{}$ and a hybrid controller~$(\kappa,\Vscr,\jset[c],\fmap[c],\jmap[c])$ that is synergistic relative to $\Ascr$ for~\eqref{eq:sysx} with robustness margin $\Omega$, we say that it has synergy gap exceeding $\delta$ if, for each $\theta\in\Omega$ and each $(x,\xc)\in\Psi[\theta]\minus\Amc[\theta]$,
$\delta(x,\xc)<\gap{\V[\theta]}(x,\xc).$

\begin{theorem}\label{thm:rgas}
Suppose that Assumption~\ref{ass:data} holds. Given a compact set $\Omega$, a positive function $\delta:\Xmc\x\Xmc[c]\to\R{}$, and a collection of compact subsets $\Ascr\ceq\{\Amc[\theta]\}_{\theta\in\Omega}$ of $\Xmc\x\Xmc[c]$, if~$(\kappa,\Vscr,\jset[c],\fmap[c],\jmap[c])$ is synergistic relative to $\Ascr$ for~\eqref{eq:sysx} with robustness margin $\Omega$ and synergy gap exceeding $\delta$,
then, for each $\theta\in\Omega$, the set $\Amc[\theta]$ is globally pre-asymptotically stable for~\eqref{eq:rclosed}. If, for each $(x,\xc)\in\fset[\Omega]\minus\jset[\Omega]$,~\ref{ass:VC2} is satisfied, then $\Amc[\theta]$ is globally asymptotically stable for~\eqref{eq:rclosed}.
\end{theorem}
\begin{proof}
For each $\theta\in\Omega$, it follows from~\ref{ass:V} that each sublevel set of $\V[\theta]$ is compact and, since it is also forward pre-invariant as shown in Lemma~\ref{lem:rstability}, we have that each solution to~\eqref{eq:rclosed} is bounded.
In addition, it follows from the proof of Lemma~\ref{lem:rstability} that the growth of $\V[\theta]$ along jumps of~\eqref{eq:rclosed} is bounded by~\eqref{eq:udtt} and, since $\delta(x,\xc)>0$ by assumption, it follows from~\cite[Theorem~8.2]{goebel_hybrid_2012} that each complete solution to~\eqref{eq:rclosed} approaches the largest weakly invariant subset of 
$\V[\theta]\inv(r)\cap\cl{\uc[theta]\inv(0)}$
for some $r$ in the image of $\V[\theta]$, which is to say that each complete solution to~\eqref{eq:rclosed} approaches $\Psi[\theta]\cap \fset[\Omega]$. Since each point $(x,\xc)\in\Psi[\theta]\minus\Amc[\theta]$ belongs to $\jset[\Omega]\minus\fset[\Omega]$ by Assumption~\ref{ass:Psi}, it follows that each complete solution to~\eqref{eq:rclosed} converges to $\Amc[\theta]$, which concludes the proof of global pre-attractivity of $\Amc[\theta]$ for~\eqref{eq:rclosed}. 
The proof of stability of $\Amc[\theta]$ for~\eqref{eq:closed} follows closely the proof of Lemma~\ref{lem:stability}. We conclude that $\Amc[\theta]$ is globally pre-asymptotically stable for~\eqref{eq:rclosed}. Global asymptotic stability of $\Amc[\theta]$ for~\eqref{eq:rclosed} under assumption~\ref{ass:VC2} follows directly from global pre-asymptotic stability and completeness of solutions, as shown in Lemma~\ref{lem:rstability}.
\end{proof}



In the next section, we apply the proposed controller to the design of adaptive synergistic feedback control laws for a class of affine systems with matched uncertainties.

\section{Adaptive Backstepping of Synergistic Hybrid Feedback for Affine Control Systems}\label{sec:adaptive}
\subsection{Nominal Synergistic Hybrid Feedback}\label{sec:given}
In this section, we apply the controller design of Section~\ref{sec:robust} to the problem of global asymptotic stabilization of a compact set $\Amc\subset\Xmc\x\Xmc[c]$ for a control affine system subject to parametric uncertainty, where $\Xmc$ and $\Xmc[c]$ denote the spaces of the state and controller variables, respectively. In this direction, let $\fmap[\theta]$ in~\eqref{eq:sysx} be given by
\begin{equation}
    \label{eq:sysz}
    \fmap[\theta](x,\xc,u)\ceq\psi[x](x,\xc)+\psi[u](x,\xc)u+\psi[theta](x,\xc)\theta
\end{equation}
for each $(x,\xc,u)\in\Xmc\x\Xmc[c]\x\Umc$, where $u$ denotes an input variable subject to the constraint $u\in\Umc$, and 
\begin{equation}
    \label{eq:Omega}
    \theta\in\Omega\ceq\{\theta\in\R \ell:\norm{\theta}\leq \tz\}
\end{equation}
represents the parametric uncertainty of the model whose norm is assumed to be bounded by a known parameter $\theta_0\in\Rnneg$. The controller design in this section is applicable under the assumption of matched uncertainties stated next.

\begin{assumption}\label{ass:affine}
There exists a continuously differentiable function $\psi[th]$ such that $\psi[theta](x,\xc)=\psi[u](x,\xc)\psi[th](x,\xc)$ for each $(x,\xc)\in\Xmc\x\Xmc[c]$.
\end{assumption}

In addition, we assume that we are given a synergistic hybrid controller for the nominal (unperturbed) system as defined next.

\begin{definition}\label{def:Hc}
Given a compact set $\Amc\subset\Xmc\x\Xmc[c]$ and a continuous function $\delta:\Xmc\x\Xmc[c]\to\R{}$, the hybrid controller~$(\kappa[0],\V[0],\jset[c],\fmap[c])$ is said to be \emph{nominally} synergistic relative to $\Amc$ for~\eqref{eq:sysz} with synergy gap exceeding $\delta$ if it is synergistic relative to $\Amc$ for
\begin{equation}
    \label{eq:unperturbed}
    \dot x=\fmap[0](x,\xc,u)\ceq\psi[x](x,\xc)+\psi[u](x,\xc)u
\end{equation}
with synergy gap exceeding $\delta$, and $\V[0]$ is continuously differentiable on $\{(x,\xc)\in\Xmc\x\Xmc[c]:\V[0](x,\xc)<+\infty.\}$.
\end{definition}

The dynamical system~\eqref{eq:unperturbed} is obtained from~\eqref{eq:sysz} by considering that there are no perturbations, i.e., $\theta=0$. It follows from Theorem~\ref{thm:gas} that $\Amc$ is globally asymptotically stable for the closed-loop system $\Hmc$ in~\eqref{eq:closed} resulting from the interconnection of~\eqref{eq:unperturbed} and a nominally synergistic controller relative to $\Amc$ for~\eqref{eq:sysz} when $\theta=0$.
In the next section, we present variations of the nominal synergistic controller so as to deal with nonzero disturbances. 

\subsection{Adaptive Synergistic Hybrid Feedback}\label{sec:start}

In this section, we modify the nominal synergistic controller given in Section~\ref{sec:given} to globally asymptotically stabilize \begin{equation}
    \label{eq:Amc0}
    \Amc[1,\theta]\ceq\Amc\x\{\theta\}
    \end{equation}
for the closed-loop system when $\theta$ in~\eqref{eq:sysz} is nonzero.%
\footnote{As the controller design exploits ideas in the literature of adaptive control, we refer the reader to~\cite{zhou_adaptive_2008} for an overview of adaptive controller design and backstepping under the influence of model uncertainty.}
In this direction, let $\th\in\R\ell$ denote an estimate of the parameter $\theta$ that is generated via
\begin{equation}
\label{eq:thdot}
\olddot{\th}=\Gamma[1]\Proj(\psi[theta](x,\xc)\tp\grad[x]\V[0](x,\xc),\th),
\end{equation}
where $\Gamma[1]\in\R{\ell\x\ell}$ is a positive definite matrix and $\Proj:\R\ell\x\R \ell\to\R\ell$ is given by
\begin{equation}
\label{eq:Proj}
\Proj(\eta,\th)\ceq\begin{cases}
\begin{minipage}{0.33\textwidth}
$\eta$ \hfill if  $\p(\th)\leq 0$  or $\grad\p(\th)\tp\eta\leq 0$
\end{minipage}\\
\begin{minipage}{0.33\textwidth}
$\left(\eye{\ell}-\frac{p(\th)\grad\p(\th)\grad\p(\th)\tp}{\grad\p(\th)\tp\grad\p(\th)}\right)\eta$ \hfill otherwise
\end{minipage}
\end{cases}
\end{equation}
for each $(\eta,\th)\in\R\ell\x\R\ell$,
\begin{equation}
\label{eq:p}
\p(\th)\ceq\frac{\th\tp\th-\tz^2}{\slack^2+2\slack\tz}
\end{equation}
for each $\th\in\R\ell$, with $\slack>0$ and $\tz>0$ given in~\eqref{eq:Omega}, and $\psi[theta]$ as in~\eqref{eq:sysz}. The function $\Proj$ in~\eqref{eq:Proj} has the following properties (cf.~\cite{cai_sufficiently_2005}): 
\begin{enumerate}[label=(P\arabic*)]
\item\label{ass:ProjLip} $\Proj$ is Lipschitz continuous;
\item\label{ass:thbnd} Each solution $t\mapsto\th(t)$ to
$\olddot{\th}=\Gamma[1]\Proj(\eta(t),\th),$
from $\th\in\Omega+\slack\cl{\ball}$ with input $t\mapsto\eta(t)$ satisfies $\rge\th\subset\Omega+\slack\cl{\ball}$;
\item\label{ass:tildebnd} Given $\theta\in\Omega$,
$(\theta-\th)\tp\Proj(\eta,\th)\geq(\theta-\th)\tp\eta$ for each $(\eta,\th)\in\R\ell\x\R\ell$;
\end{enumerate}
with $\slack>0$ as in~\eqref{eq:p}. Given a hybrid controller~$(\kappa[0],\V[0],\jset[c],\fmap[c])$ that is nominally synergistic relative to $\Amc$ for~\eqref{eq:sysz} with synergy gap exceeding $\delta$, and the controller variable $\xc[1]\ceq(\xc,\th)\in\Xmc[c,1]\ceq\Xmc[c]\x(\Omega+\slack\cl{\ball})$, we define %
\begin{subequations}\label{eq:data0}
\begin{align}
\label{eq:kappa0}
\kappa[1](x,\xc[1])&\ceq\kappa[0](x,\xc)-\psi[th](x,\xc)\th\\
\label{eq:V0}
\V[1,\theta](x,\xc[1])&\ceq\V[0](x,\xc)+\frac{1}{2}(\theta-\th)\tp\Gamma[1]\inv(\theta-\th)\\
\label{eq:jset0} \jset[c,1](x,\xc[1])&\ceq\jset[c](x,\xc)\x(\Omega+\slack\cl{\ball})\\
\label{eq:fmapc0}\fmap[c,1](x,\xc[1]) &= \bmtx{\fmap[c](x,\xc)\\ \Gamma[1]\Proj(\psi[theta](x,\xc)\tp\grad[x]\V[0](x,\xc),\th)}
\end{align}
\end{subequations}
for each $(x,\xc[1])\in\Xmc\x\Xmc[c,1]$, where $\psi[th]$ comes from Assumption~\ref{ass:affine}. 
The hybrid closed-loop system resulting from the interconnection between~\eqref{eq:sysz} and the hybrid controller~$(\kappa[1],\V[1,\theta],\jset[c,1],\fmap[c,1])$, is given by
\begin{subequations}
\mathtoolsset{showonlyrefs=false}
\label{eq:sys0}
\begin{align}
(\dot x,\dot{xc1})&\in\fmap[cl,1](x,\xc[1]) & (x,\xc[1])&\in\fset[1]\\
(x\pl, \xc[1]\pl)&\in\jmap[cl,1](x,\xc[1]) & (x,\xc[1])&\in\jset[1]
\end{align}
\end{subequations}
where 
\begin{align*}
\fset[1]&\ceq\{(x,\xc[1])\in\Xmc\x\Xmc[c,1]:\gap{\V[1,\theta]}(x,\xc[1])\leq\delta(x,\xc)\}\\
\jset[1]&\ceq\{(x,\xc[1])\in\Xmc\x\Xmc[c,1]:\gap{\V[1,\theta]}(x,\xc[1])\geq\delta(x,\xc)\}
\end{align*}
and
\begin{subequations}
\mathtoolsset{showonlyrefs=false}
\begin{align}
\label{eq:fmap0}
&\begin{multlined}[0.45\textwidth]
\fmap[cl,1](x,\xc[1])\ceq\bmtx{\fmap[\theta](x,\xc,\kappa[1](x,\xc[1]))\\ 
    \fmap[c,1](x,\xc[1])} \\ \forall (x,\xc[1])\in\fset[1]
\end{multlined}\\
&\begin{minipage}{0.45\textwidth}
$\jmap[cl,1](x,\xc[1])\ceq\bmtx{x\\ \qminV{\V[1,\theta]}(x,\xc[1])} \hfill \forall (x,\xc[1])\in\jset[1].$
\end{minipage}
\end{align}
\end{subequations}
where, for each $(x,\xc[1])\in\Xmc\x\Xmc[c,1]$,
\begin{subequations}\label{eq:minstt}
    \mathtoolsset{showonlyrefs=false}
    \begin{align}
    \label{eq:minVtt}\minV{\V[1,\theta]}(x,\xc[1])&=\minV{\V[0]}(x,\xc)\\
    \label{eq:qminVtt}\qminV{\V[1,\theta]}(x,\xc[1])%
    &=\qminV{\V[0]}(x,\xc)\x\{\theta\}\\
    \label{eq:gaptt0}\gap{\V[1,\theta]}(x,\xc[1]) &= \gap{\V[0]}(x,\xc)+\frac{1}{2}(\theta-\th)\tp\Gamma[1]\inv(\theta-\th)
    \end{align}
    \end{subequations}
are directly computed from~\eqref{eq:minV},~\eqref{eq:qminV} and~\eqref{eq:gap}, respectively.
\begin{remark}\label{rem:nonworking}
For the hybrid controller~$(\kappa[1],\V[1,\theta],\jset[c,1],\fmap[c,1])$, the functions~\eqref{eq:minstt} are not realizable, because $\gap{\V[1,\theta]}$ and $\qminV{\V[1,\theta]}$ in~\eqref{eq:minstt} depend on the unknown constant $\theta$. This dependence will be removed when we show that there exists $\jmap[c,1]:\Xmc\x\Xmc[c,1]\tto\Xmc[c,1]$ such that the hybrid controller~$(\kappa[1],\Vscr[1],\jset[c,1],\fmap[c,1],\jmap[c,1])$ with $\Vscr\ceq\{\V[1,\theta]\}_{\theta\in\Omega}$ is synergistic relative to $\Ascr[1]\ceq\{\Amc[1,\theta]\}_{\theta\in\Omega}$ for~\eqref{eq:sysz} with robustness margin $\Omega$.
\end{remark}

To design~\eqref{eq:rsysq}, we start by showing that the hybrid controller~$(\kappa[1],\V[1,\theta],\jset[c,1],\fmap[c,1])$ is synergistic relative to $\Amc[1,\theta]$ for~\eqref{eq:sysz}.

\begin{proposition}\label{pro:candidate}
Suppose that the sets $\Xmc$, $\Xmc[c]$, $\Umc$, and the set-valued map $\fmap[\theta]$ in~\eqref{eq:sysz} satisfy Assumption~\ref{ass:data}, and that Assumption~\ref{ass:affine} holds. Given $\theta\in\Omega$, a compact set $\Amc\subset\Xmc\x\Xmc[c]$, and a hybrid controller~$(\kappa[0],\V[0],\jset[c],\fmap[c])$ that is nominally synergistic relative to $\Amc$ for~\eqref{eq:sysz}, the controller~$(\kappa[1],\V[1,\theta],\jset[c,1],\fmap[c,1])$ given in~\eqref{eq:data0} is a synergistic candidate relative to $\Amc[1,\theta]$ for~\eqref{eq:sysz}.
\end{proposition}
\begin{proof}
The optimization problems in~\eqref{eq:mins} are feasible for each $x\in\Xmc$, because they are feasible for $\V[0]$, hence~\ref{ass:Qbasic} is satisfied.

Since $\V[1,\theta]$ corresponds to the sum of $\V[0]$ with $(\theta-\th)\tp\Gamma[1]\inv(\theta-\th)/2$ and both terms are continuous, it follows that $\V[1,\theta]$ is continuous. 
Since $\V[0]$ is positive definite with respect to $\Amc$ and $\th\mapsto (\theta-\th)\tp\Gamma[1]\inv(\theta-\th)$ is positive definite relative to $\theta$, it follows that $\V[1,\theta]$ is positive definite relative to $\Amc[1,\theta]$. It follows from the assumption that $\V[0]\inv([0,c])$ is compact for each $c\geq 0$ and radial unboundedness of $\th\mapsto (\theta-\th)\tp\Gamma[1]\inv(\theta-\th)$ relative to $\theta$ that $\V[1,\theta]\inv([0,c])$ is compact for each $c\geq0$, thus proving that $\V[1,\theta]$ satisfies~\ref{ass:V}.

From~\eqref{eq:jset0}, we have that $\jset[c,1](x, \xc[1])$ is the Cartesian product between $\jset[c](x,\xc)$ and $\Omega+\slack\cl{\ball}$ for each $(x,\xc)\in\Xmc\x\Xmc[c]$. Since $\jset[c]$ satisfies~\ref{ass:Q} by assumption, we have that $\jset[c,1]$ also satisfies~\ref{ass:Q}. Since $\kappa[0]$ satisfies~\ref{ass:kappa}, then $\kappa[1]$ also satisfies~\ref{ass:kappa}.
\end{proof}

\begin{proposition}\label{pro:uc0}
    Suppose that the sets $\Xmc$, $\Xmc[c]$, $\Umc$, and the set-valued map $\fmap[\theta]$ in~\eqref{eq:sysz} satisfy Assumption~\ref{ass:data}, and that Assumption~\ref{ass:affine} holds. Given $\theta\in\Omega$, a compact set $\Amc\subset\Xmc\x\Xmc[c]$, and a hybrid controller~$(\kappa[0],\V[0],\jset[c],\fmap[c])$ that is nominally synergistic relative to $\Amc$ for~\eqref{eq:sysz}, the controller~$(\kappa[1],\V[1,\theta],\jset[c,1],\fmap[c,1])$ given in~\eqref{eq:data0} satisfies~\ref{ass:uc}.
\end{proposition}
\begin{proof}
It follows from~\eqref{eq:V0},~\eqref{eq:fmap0} and~\ref{ass:tildebnd} that, for each $(x,\xc[1])\in\Xmc\x\Xmc[c,1]$ and each $f_{cl,1}\in\fmap[cl,1](x,\xc[1])$
\begin{equation*}\hspace*{-2.2pt}
\begin{aligned}\label{eq:DV0_1}
\grad\V[1,\theta](x,\xc[1])\tp f_{cl,1}\leq&\grad\V[0](x,\xc)\tp\!\bmtx{\fmap[\theta](x,\xc,\kappa[1](x,\xc[1]))\\ f_c}\\
&-(\theta-\th)\tp\psi[theta](x,\xc)\tp\grad[x]\V[0](x,\xc)
\end{aligned}
\end{equation*}
where $f_c\in\fmap[c](x,\xc)$ is the component of $f_{cl,1}$ that determines the dynamics of $\xc$, i.e., $\dot{xc}=f_c$.
Replacing~\eqref{eq:sysz} and~\eqref{eq:kappa0} in~\eqref{eq:DV0_1}, 
\IfArxiv{we obtain
\begin{xequation}\label{eq:DV0_2}
\begin{aligned}
\grad&\V[1,\theta](x,\xc[1])\tp f_{cl,1}\leq\grad\V[0](x,\xc)\tp \bmtx{\fmap[0](x,\xc,\kappa[0](x,\xc))\\ f_c}\\
&+\grad[x]\V[0](x,\xc)\tp(-\psi[u](x,\xc)\psi[th](x,\xc)\th+\psi[theta](x,\xc)\theta)\\
&-(\theta-\th)\tp\psi[theta](x,\xc)\tp\grad\V[0](x,\xc)
\end{aligned}
\end{xequation}
for each $(x,\xc[1])\in\Xmc\x\Xmc[c,1]$  with $\fmap[0]$ given in~\eqref{eq:unperturbed}. Hence, }{}
it follows from~Assumption~\ref{ass:affine} that
\begin{equation*}
\grad\V[1,\theta](x,\xc[1])\tp f_{cl,1}\leq\grad\V[0](x,\xc)\tp \bmtx{\fmap[0](x,\xc,\kappa[0](x,\xc))\\ f_c}
\end{equation*}
for each $(x,\xc[1])\in\Xmc\x\Xmc[c,1]$. From the assumption that the hybrid controller~\eqref{eq:sysq} with data $(\kappa[0],\V[0],\jset[c],\fmap[c])$ is synergistic relative to $\Amc$ for~\eqref{eq:unperturbed}, we have that 
%
$\grad\V[1,\theta](x,\xc[1])\tp f_{cl,1}\leq 0$
%
for each $(x,\xc[1])\in\Xmc\x\Xmc[c,1]$ and each $f_{cl,1}\in\fmap[cl,1](x,\xc[1])$, which proves~\ref{ass:uc}.
\end{proof}

Since the hybrid controller~$(\kappa[1],\V[1,\theta],\jset[c,1],\fmap[c,1])$ satisfies~\ref{ass:uc}, we have that $\V[1,\theta]$ is nonincreasing along solutions to the closed-loop system~\eqref{eq:sys0}, but satisfying~\ref{ass:Psi} requires further assumptions on the data, as shown next.

\begin{proposition}\label{pro:synergy0}
    Suppose that the sets $\Xmc$, $\Xmc[c]$, $\Umc$, and the set-valued map $\fmap[\theta]$ in~\eqref{eq:sysz} satisfy Assumption~\ref{ass:data}, and that Assumption~\ref{ass:affine} holds. Given $\theta\in\Omega$, a compact set $\Amc\subset\Xmc\x\Xmc[c]$, and a hybrid controller~$(\kappa[0],\V[0],\jset[c],\fmap[c])$ that is nominally synergistic relative to $\Amc$ for~\eqref{eq:sysz} with synergy gap exceeding $\delta$,  let $\Psi$ denote the largest weakly invariant subset of
\begin{align*}\label{eq:Psi}
(\dot{x},\dot{xc})\in \fmap[cl,0](x,\xc)=\pmtx{\fmap[0](x,\xc,\kappa[0](x,\xc))\\ \fmap[c](x,\xc)}
\end{align*}
on $(x,\xc)\in\Emc\ceq\{(x,\xc)\in\Xmc\x\Xmc[c]:\grad\V[0](x,\xc)\tp f_{cl,0}=0\\
\text{ for some }f_{cl,0}\in\fmap[cl,0](x,\xc)\}$ 
and let $\Psi[1,\theta]$ denote the largest weakly invariant subset of
\begin{align*}
(\dot{x},\dot{xc1})&\in\fmap[cl,1](x,\xc[1]) & (x,\xc[1])&\in\Emc[1]
\end{align*}
with $\Emc[1]\ceq\{(x,\xc[1])\in\Xmc\x\Xmc[c,1]:\grad\V[1,\theta](x,\xc[1])\tp f_{cl,1}=0\
\text{ for some }f_{cl,1}\in\fmap[cl,1](x,\xc[1])\}$.
If the projection of $\Psi[1,\theta]\minus\Amc[1,\theta]$ onto $\Xmc\x\Xmc[c]$ is a subset of $\Psi\minus\Amc$, i.e.,
\footnote{Given a subset $S$ of $X\ceq X_1\x X_2$, the projection of $S$ onto $X_1$ is represented by 
$\pi[X_1](S)\ceq\{x_1\in X_1:(x_1,x_2)\in S\text{ for some }x_2\in X_2\}.$
Similarly, the projection of $S$ onto $X_2$ is denoted by
$\pi[X_2](S)\ceq\{x_2\in X_2:(x_1,x_2)\in S\text{ for some }x_1\in X_1\}.$ }
\begin{equation}
\label{eq:level}  \pi[{\Xmc\x\Xmc[c]}](\Psi[1,\theta]\minus\Amc[1,\theta])\subset\Psi\minus\Amc,
\end{equation}
then the hybrid controller~$(\kappa[1],\V[1,\theta],\jset[c,1],\fmap[c,1])$ in~\eqref{eq:data0} is synergistic relative to $\Amc[1,\theta]$ for~\eqref{eq:sysz} with synergy gap exceeding $\delta$.
\end{proposition}

\begin{proof}
It follows from the definition of $\gap{\V[1,\theta]}$ in~\eqref{eq:gaptt0} that $\gap{\V[1,\theta]}(x,\xc[1])$ is the sum of $\gap{\V[0]}(x,\xc)$ with a quadratic nonnegative term, hence
\begin{equation}\label{eq:gapttVsGap}
\gap{\V[1,\theta]}(x,\xc[1])\geq\gap{{\V[0]}}(x,\xc)
\end{equation}
for each $(x,\xc[1])\in\Psi[1,\theta]\minus\Amc[1,\theta]$ and, consequently, we have that
\begin{equation}\label{eq:deltabnd_1}
\begin{aligned}
\deltabnd[2]\ceq&\inf\{\gap{\V[1,\theta]}(x,\xc,\theta):(x,\xc[1])\in\Psi[1,\theta]\minus\Amc[1,\theta]\}\\
\geq&\inf\left\{\gap{{\V[0]}}(x,\xc):(x,\xc[1])\in\Psi[1,\theta]\minus\Amc[1,\theta]\right\}.
\end{aligned}
\end{equation}
The fact that $(x,\xc[1])\in\Psi[1,\theta]\minus\Amc[1,\theta]$ implies $(x,\xc)\in\pi[{\Xmc\x\Xmc[c]}](\Psi[1,\theta]\minus\Amc[1,\theta])$ together with~\eqref{eq:deltabnd_1} allow us to derive the following inequality:
    $\deltabnd[2]\geq\inf\left\{\gap{{\V[0]}}(x,\xc):(x,\xc)\in\pi[{\Xmc\x\Xmc[c]}](\Psi[1,\theta]\minus\Amc[1,\theta])\right\}.$
It follows 
from~\eqref{eq:level} that 
$\deltabnd[2]\geq\inf\{\gap{{\V[0]}}(x,\xc):(x,\xc)\in\Psi\minus\Amc\}$
which is greater than zero by the assumption that the controller~$(\kappa[0],\V[0],\jset[c],\fmap[c])$ is synergistic relative to $\Amc$ for~\eqref{eq:unperturbed} with synergy gap exceeding $\delta$. In addition, we have that 
$\gap{\V[1,\theta]}(x,\xc[1])\geq\gap{{\V[0]}}(x,\xc)>\delta(x,\xc)$
for each $(x,\xc[1])\in\Psi[1,\theta]\minus\Amc[1,\theta]$, which proves that the hybrid controller~$(\kappa[1],\V[1,\theta],\jset[c,1],\fmap[c,1])$ in~\eqref{eq:data0} is synergistic relative to $\Amc[1,\theta]$ for~\eqref{eq:sysz} with synergy gap exceeding $\delta$.
\end{proof}

In the next result, we complete the construction of the robust synergistic controller~\eqref{eq:rsysq} from the data of a nominally synergistic controller $(\kappa[0],\V[0],\jset[c],\fmap[c])$, by designing a set-valued map $\jmap[c,1]:\Xmc\x\Xmc[c,1]\tto\Xmc$ that is outer semicontinuous, locally bounded and satisfies~\eqref{eq:Gtt}.

\begin{proposition}\label{pro:robust0}
    Suppose that the sets $\Xmc$, $\Xmc[c]$, $\Umc$, and the set-valued map $\fmap[\theta]$ in~\eqref{eq:sysz} satisfy Assumption~\ref{ass:data}, and that Assumption~\ref{ass:affine} holds. Given $\Omega$ in~\eqref{eq:Omega}, a compact set $\Amc\subset\Xmc\x\Xmc[c]$, and a hybrid controller~$(\kappa[0],\V[0],\jset[c],\fmap[c])$ that is nominally synergistic relative to $\Amc$ for~\eqref{eq:sysz} with synergy gap exceeding $\delta$, $\Ascr[1]\ceq\{\Amc[1,\theta]\}_{\theta\in\Omega}$ with $\Amc[1,\theta]$ in~\eqref{eq:Amc0}, $\Vscr[1]\ceq\{\V[1,\theta]\}_{\theta\in\Omega}$ with $\V[1,\theta]$ in~\eqref{eq:V0}, then the hybrid controller~$(\kappa[1],\Vscr[1],\jset[c,1],\fmap[c,1], \jmap[c,1])$ where 
\begin{equation}\label{eq:jmapc0}
    \jmap[c,1](x,\xc[1])\ceq\qminV{{\V[0]}}(x,\xc)\x\hat{\jmap}(\th)
\end{equation} 
for each $(x,\xc[1])\in\Xmc\x\Xmc[c,1]$, and
\begin{equation*}
\begin{multlined}
\hat{\jmap}(\th)\ceq\argmax_{g\in\Omega+\slack\cl{\ball}}\ \min_{\theta\in\Omega}\ (\theta-\th)\tp\Gamma[1]\inv(\theta-\th)\\
-(\theta-g)\tp\Gamma[1]\inv(\theta-g)
\end{multlined}
\end{equation*}
for each $\th\in\Omega+\slack\cl{\ball}$, is synergistic relative to $\Ascr[1]$ for~\eqref{eq:sysz} with robustness margin $\Omega$ and synergy gap exceeding $\delta$.
\end{proposition}
\begin{proof}
In Proposition~\ref{pro:synergy0} we demonstrate that the hybrid controller~$(\kappa[1],\V[1,\theta],\jset[c,1],\fmap[c,1])$ is synergistic relative to $\Amc[1,\theta]$ as required by Definition~\ref{def:robust}. It remains to be shown that the hybrid controller~$(\kappa[1],\Vscr[1],\jset[c,1],\fmap[c,1], \jmap[c,1])$ satisfies assumptions~\ref{ass:Vtheta},~\ref{ass:Gc_regular} and~\ref{ass:Gc_decreases}.
To prove~\ref{ass:Vtheta}, one must show that $\Xmc\x\Xmc[c,1]\subset \dom\V[1,\theta]$. From the definition of $\V[1,\theta]$ in~\eqref{eq:V0}, we have that $\dom\V[1,\theta]=\dom\V[0]\x (\Omega+\epsilon\cl{\ball})$. It follows from the assumption that the hybrid controller~$(\kappa[0],\V[0],\jset[c],\fmap[c])$ is nominally synergistic relative to $\Amc$ for~\eqref{eq:sysz} that $\Xmc\x\Xmc[c]\subset \dom\V[0]$, hence $\Xmc\x\Xmc[c,1]\subset\dom\V[1,\theta]$. The function $(x,\xc[1],\theta)\mapsto \V[1](x,\xc[1],\theta)\ceq\V[1,\theta](x,\xc[1])$ is continuous because it results from the composition of continuous functions, hence~\ref{ass:Vtheta} holds.

To prove~\ref{ass:Gc_regular} and~\ref{ass:Gc_decreases}, one must show that $\jmap[c,1]$ is outer semicontinuous, locally bounded and that it satisfies~\eqref{eq:Gtt}.
Since $\jmap[c,1](x,\xc[1])$ is the Cartesian product of $\qminV{\V[0]}(x, \xc)$ and $\hat{\jmap}(\th)$ for each $(x,\xc[1])\in\Xmc\x\Xmc[c,1]$ and $\qminV{\V[0]}$  is outer semicontinuous and locally bounded as proved in Lemma~\ref{lem:regular}, to demonstrate that~\ref{ass:Gc_regular} is satisfied it only remains to be shown that $\hat{\jmap}$ is outer semicontinuous and locally bounded.
Let
$h(g,\theta)\ceq (\theta-\th)\tp\Gamma[1]\inv(\theta-\th)-(\theta-g)\tp\Gamma[1]\inv(\theta-g)$
for each $(g,\theta)\in(\Omega+\slack\cl{\ball})\x\Omega$. Since $h$ results from the composition of continuous functions it is also continuous. It follows from the compactness of $\Omega$ and from~\cite[Theorem~9.14]{Sundaram1996} that 
\begin{align}
\label{eq:lbarh}
\lbar{h}(g)&\ceq\min\{h(g,\theta):\theta\in\Omega\}&
 &\forall g\in\Omega+\slack\cl{\ball}
\end{align}
is continuous. Since $\jset[c,1]$ is continuous and compact-valued, it follows from the continuity of~\eqref{eq:lbarh} and~\cite[Theorem~9.14]{Sundaram1996} that $\hat{\jmap}$ is compact-valued and upper semicontinuous. The remainder of the proof of outer semicontinuity and local boundedness of $\hat{\jmap}$ follows closely that of Lemma~\ref{lem:regular}, thus it will be omitted.

The fact that $\jmap[c,1]$ satisfies~\eqref{eq:Gtt} follows from the observations in Remark~\ref{rem:Gtt} by noticing that $\Omega$ and $\Omega+\slack\cl{\ball}$ are convex and compact spaces and the function $h$, which can be rewritten as
$h(g,\theta)=2\theta\tp\Gamma[1]\inv(g-\th)-g\tp\Gamma[1]\inv g+\th\tp \Gamma[1]\inv\th$
for each $(g,\theta)\in(\Omega+\slack\cl{\ball})\x\Omega$, is quasi-concave as a function of $g$ and quasi-convex as a function of $\theta$.
\end{proof}

The hybrid closed-loop system resulting from the interconnection between~$(\kappa[1],\Vscr[1],\jset[c,1],\fmap[c,1],\jmap[c,1])$ and~\eqref{eq:sysz} is given by:
\begin{subequations}
\mathtoolsset{showonlyrefs=false}
\label{eq:sys0Omega}
\begin{align}
(\dot x,\dot{xc1})&\in\fmap[cl,1](x,\xc[1]) & (x,\xc[1])&\in\fset[\Omega,1]\\
(x\pl, \xc[1]\pl)&\in\jmap[\Omega,1](x,\xc[1]) & (x,\xc[1])&\in\jset[\Omega,1]
\end{align}
\end{subequations}
where \par
\resizebox{0.485\textwidth}{!}{\hspace*{-10.2pt}\parbox{0.5\textwidth}{\begin{align*}
\fset[\Omega,1]&\ceq\left\{(x,\xc[1])\in\Xmc\x\Xmc[c,1]:\min_{\theta\in\Omega}\gap{\V[1,\theta]}(x,\xc[1])\leq\delta(x,\xc)\right\}\\
\jset[\Omega,1]&\ceq\left\{(x,\xc[1])\in\Xmc\x\Xmc[c,1]:\min_{\theta\in\Omega}\gap{\V[1,\theta]}(x,\xc[1])\geq\delta(x,\xc)\right\}
\end{align*}}}
and
\begin{align}
\jmap[\Omega,1](x,\xc[1])\ceq\bmtx{x\\ \jmap[c,1](x,\xc[1])} \hfill \forall (x,\xc[1])\in\jset[\Omega,1].
\end{align}

Global asymptotic stability of $\Amc[1,\theta]$ for~\eqref{eq:sys0Omega} follows from the application of Theorem~\ref{thm:rgas} and it is summarized in the next corollary.

\begin{corollary}\label{cor:gas0}
    Suppose that the sets $\Xmc$, $\Xmc[c]$, $\Umc$, and the set-valued map $\fmap[\theta]$ in~\eqref{eq:sysz} satisfy Assumption~\ref{ass:data}, and that Assumption~\ref{ass:affine} holds. Given $\Omega$ in~\eqref{eq:Omega}, a positive function $\delta:\Xmc\x\Xmc[c]\mapsto\R{}$, a compact set $\Amc\subset\Xmc\x\Xmc[c]$, and a hybrid controller~$(\kappa[0],\V[0],\jset[c],\fmap[c])$ that is nominally synergistic relative to $\Amc$ for~\eqref{eq:sysz} with synergy gap exceeding $\delta$, for each $\theta\in\Omega$, the set $\Amc[1,\theta]$ is globally asymptotically stable for~\eqref{eq:sys0Omega}.
\end{corollary}
\begin{proof}
It follows from~\eqref{eq:gapttVsGap} that 
$\min\{\gap{\V[1,\theta]}(x,\xc[1]):\theta\in\Omega\}\geq\gap{\V[0]}(x,\xc)$
for each $(x,\xc[1])\in\Psi[1,\theta]\minus\Amc[1,\theta]$. Since $\gap{\V[0]}(x,\xc)>\delta(x,\xc)$ for each $(x,\xc[1])\in\Psi[1,\theta]\minus\Amc[1,\theta]$ as shown in the proof of Proposition~\ref{pro:synergy0}, and $\delta$ satisfies~\ref{ass:deltazero}, the conditions of Theorem~\ref{thm:rgas} apply and we are able to conclude that $\Amc[1,\theta]$ is globally asymptotically stable for~\eqref{eq:sys0Omega}.
\end{proof}

\subsection{Backstepping}\label{sec:back}

Given a nominally synergistic controller~$(\kappa[0],\V[0],\jset[c],\fmap[c])$, we extend the dynamics of the controller in Section~\ref{sec:start} to include the input $u$ as a controller state:\footnote{Alternatively, one may consider $u$ as a plant state rather than a controller state, in which case $u$ would remain constant during jumps. We have included $u$ as a controller variable because it is an approach less often found in the literature.}
\begin{equation}\label{eq:systt1}
\begin{aligned}
&\dot{xc2}\in\fmap[c,2](x,\xc[2])\\
&\ceq\left\{\bmtx{%
f_c\\
\Gamma[1]\Proj(\ff(x,\xc[2]),\th)\\
f_u(x,\xc[2])+\D[\xc](\kappa[1](x,\xc[1]))f_c}:f_c\in\fmap[c](x,\xc)\right\}
\end{aligned}
\end{equation}
with $\xc[2]\ceq(\xc[1],u)\in\Xmc[c,2]\ceq\Xmc[c]\x(\Omega+\slack\cl{\ball})\x\R m$, $\Gamma[2]\in\R{m\x m}$ positive definite, $\ku>0$, 
\begin{multline}\label{eq:ff}
\ff(x,\xc[2])\ceq\psi[theta](x,\xc)\tp\grad[x]\V[0](x,\xc)\\
-\psi[theta](x,\xc)\tp\D[x](\kappa[1](x,\xc[1]))\tp\Gamma[2]\inv(u-\kappa[1](x,\xc[1]))
\end{multline}
for each $(x,\xc[2])\in\Xmc\x\Xmc[c,2]$, and
\begin{equation}
\label{eq:kappa1}
\begin{aligned}
f_u(x,&\xc[2])\ceq-\psi[th](x,\xc)\Gamma[1]\Proj(\ff(x,\xc[2]),\th)\\
&-\ku (u-\kappa[1](x,\xc[1]))-\Gamma[2]\psi[u](x,\xc)\tp\grad[x]\V[0](x,\xc)\\
&+\D[x](\kappa[1](x,\xc[1]))\fmap(x,\xc,u,\th)
 \end{aligned}
\end{equation}
which is defined for each $(x,\xc[2])\in\Xmc\x\Xmc[c,2]$ assuming that $\kappa[0]$ is continuously differentiable and that $\fmap(x,\xc,u,\th)=\fmap[\th](x,\xc,u)$ denotes the dynamics~\eqref{eq:sysz} with $\theta$ is equal to the estimated value $\th$.

Given the compact set $\Omega$ of possible (unknown) values of $\theta$ in~\eqref{eq:Omega}, a compact set $\Amc\subset\Xmc\x\Xmc[c]$, and a nominal synergistic controller~$(\kappa[0],\V[0],\jset[c],\fmap[c])$ relative to $\Amc$ for~\eqref{eq:sysz} with synergy gap exceeding $\delta$, the main goal of this section is to design a controller of the form~\eqref{eq:rsysq} that is synergistic relative to $\Ascr[2]\ceq\{\Amc[2,\theta]\}_{\theta\in\Omega}$ for~\eqref{eq:sysz} with robustness margin $\Omega$ and synergy gap exceeding $\delta$, where
\begin{equation}
\label{eq:Amc1}
\begin{multlined}
\Amc[2,\theta]\ceq\{(x,\xc[2])\in\Xmc\x\Xmc[c,2]:(x,\xc[1])\in\Amc[1,\theta],\\ u=\kappa[1](x,\xc[1])\}.
\end{multlined}
\end{equation}

In this direction, we define the Lyapunov function
\begin{equation}
\begin{multlined}
\label{eq:V1}
\V[2,\theta](x,\xc[2])\ceq\V[1,\theta](x,\xc[1])\\
+\frac{1}{2}(u-\kappa[1](x,\xc[1]))\tp\Gamma[2]\inv(u-\kappa[1](x,\xc[1]))
\end{multlined}
\end{equation}
for each $(x,\xc[2])\in\Xmc\x\Xmc[c,2]$ and the set-valued map
\begin{multline}
 \label{eq:jset1}\jset[c,2](x, \xc[2])\ceq\{(g_{c,1},g_u)\in\Xmc[c,2]:g_{c,1}\in\jset[c,1](x, \xc[1]),\\
 g_u=\kappa[1](x,g_{c,1})\}
 \end{multline}
 for each $(x,\xc[2])\in\Xmc\x\Xmc[c,2]$.  The choice $u=\kappa[1](x,g_{c,1})$ in~\eqref{eq:jset1} may seem peculiar, but it turns out that this value minimizes~\eqref{eq:V1} with respect to $u$, hence it is suitable for the jump logic.

 From the interconnection between~\eqref{eq:sysz} and the hybrid controller~$(\kappa[2],\V[2,\theta],\jset[c,2],\fmap[c,2])$ with $\kappa[2](x,\xc[2])=u$ for each $(x,\xc[2])\in\Xmc\times\Xmc[c,2]$, we obtain the hybrid closed-loop system
\begin{subequations}
\mathtoolsset{showonlyrefs=false}
\label{eq:sys1}
\begin{align}
\begin{multlined}
(\dot x,\dot{xc2})\in\fmap[cl,2](x,\xc[2])\quad
(x,\xc[2])\in\fset[2]\\
\ceq\{(x,\xc[2])\in\Xmc\x\Xmc[c,2]:
\gap{\V[2,\theta]}(x,\xc[2])\leq\delta(x,\xc)\}
\end{multlined}\\
\begin{multlined}
(x\pl,\xc[2]\pl)\in\jmap[cl,2](x,\xc[2])\quad
(x,\xc[2])\in\jset[2]\\
\ceq\{(x,\xc[2])\in\Xmc\x\Xmc[c,2]:\gap{\V[2,\theta]}(x,\xc[2])\geq\delta(x,\xc)\}
\end{multlined}
\end{align}
\end{subequations}
where 
\begin{subequations}
\mathtoolsset{showonlyrefs=false}
\begin{align}
\label{eq:fmap1}
\fmap[cl,2](x,\xc[2])&\ceq\bmtx{ 
    \fmap[\theta](x,\xc,u)\\ 
    \fmap[c,2](x,\xc[2])} & \forall (x,\xc[2])&\in\fset[2]\\
\jmap[cl,2](x,\xc[2])&\ceq\bmtx{
    x\\ 
    \qminV{\V[2,\theta]}(x, \xc[2])} & \forall (x,\xc[2])&\in\jset[2].
\end{align}
\end{subequations}
Note that, from the definitions~\eqref{eq:qminV} and~\eqref{eq:gap}, we have the following identities for the hybrid controller~$(\kappa[2],\V[2,\theta],\jset[c,2],\fmap[c,2])$:
\begin{align*}
\qminV{\V[2,\theta]}(x, \xc[2])&=\{(g_{c,1},g_u)\in\Xmc[c,2]:g_{c,1}\in\qminV{\V[1,\theta]}(x, \xc[1]),\\
&g_u=\kappa[1](x,g_{c,1})\},\\
\gap{\V[2,\theta]}(x,\xc[2])&=\gap{\V[1,\theta]}(x,\xc[1])+\frac{1}{2}\norm{\Gamma[2]^{-\frac{1}{2}}(u-\kappa[1](x,\xc[1]))}^2
\end{align*}
for each $(x,\xc[2])\in\Xmc\x\Xmc[c,2]$,\footnote{Since $\Gamma[2]\in\R{m\x m}$ is assumed to be positive definite, $\Gamma[2]^{-\frac{1}{2}}$ exists and is unique (cf.~\cite[Section~8.5]{bernstein_matrix_2009}).}; hence, similarly to~\eqref{eq:sys0}, the closed-loop system~\eqref{eq:sys1} is impossible to implement due to dependence on $\theta$ in $\fset[2]$,~$\jset[2]$, and~$\jmap[cl,2]$, but, similarly to the controller of Section~\ref{sec:start}, this dependence will be removed with the design of a hybrid controller that is synergistic relative to $\Ascr[2]\ceq\{\Amc[2,\theta]\}_{\theta\in\Omega}$ for~\eqref{eq:sysz} with robustness margin $\Omega$ (cf. Remark~\ref{rem:nonworking}).

We are able to prove the following result using arguments similar to those of Proposition~\ref{pro:synergy0}.

\begin{proposition}\label{pro:synergy1}
Suppose that the sets $\Xmc$, $\Xmc[c]$, $\Umc$, and the set-valued map $\fmap[\theta]$ in~\eqref{eq:sysz} satisfy Assumption~\ref{ass:data}, and that Assumption~\ref{ass:affine} holds. Given $\theta\in\Omega$, a compact set $\Amc\subset\Xmc\x\Xmc[c]$, and a hybrid controller~$(\kappa[0],\V[0],\jset[c],\fmap[c])$ that is nominally synergistic relative to $\Amc$ for~\eqref{eq:sysz} with synergy gap exceeding $\delta$, if~\eqref{eq:level} is satisfied then the hybrid controller~$(\kappa[2],\V[2,\theta],\jset[c,2],\fmap[c,2])$ is synergistic relative to $\Amc[2,\theta]$ for~\eqref{eq:sysz} with synergy gap exceeding $\delta$.
\end{proposition}
\begin{proof}
Similarly to the proof of Proposition~\ref{pro:synergy0}, it is possible to show that properties~\ref{ass:Qbasic},~\ref{ass:V} and~\ref{ass:kappa} follow directly from the fact that $\Amc[2,\theta]$ is compact and from the assumption that $(\kappa[0],\V[0],\jset[c],\fmap[c])$ is synergistic relative to $\Amc$ for~\eqref{eq:unperturbed}. It follows from the continuity of $\jset[c,1]$ and $\kappa[1]$ that $\jset[c,2]$ is continuous. That $\jset[c,2]$ is compact-valued follows from compactness of $\jset[c,1]$ and continuity of $\kappa[1]$, hence~\ref{ass:Q} is satisfied.
\IfArxiv{%
It remains to be shown that properties~\ref{ass:uc} and~\ref{ass:Psi} also hold. 
It follows from~\eqref{eq:V1} and~\eqref{eq:fmap1} that
\begin{xequation}
\label{eq:DV1_1}
\begin{aligned}
\grad\V[2,\theta]&(x,\xc[2])\tp f_{cl,2}=\grad\V[0](x,\xc)\tp 
\bmtx{\fmap[\theta](x,\xc,u)\\ f_c}\\
&-(\theta-\th)\tp\Proj(\ff(x,\xc[2]),\th)\\
&+(u-\kappa[1](x,\xc[1]))\tp\Gamma[2]\inv\Bigg(f_u(x,\xc[2])\\
&-\D(\kappa[1](x,\xc[1]))\bmtx{\fmap[\theta](x,\xc,u)\\ f_c\\ \Gamma[1]\Proj(\ff(x,\xc[2]),\th)}\Bigg)
\end{aligned}
\end{xequation}
for each $(x,\xc[2])\in\Xmc\x\Xmc[c,2]$ and each $f_{cl,2}\in\fmap[cl,2](x,\xc[2])$, where $\V[2,\theta]$ is continuously differentiable and $f_c\in\fmap[c](x,\xc)$ is the component of $f_{cl,2}$ that describes the dynamics of $\xc$. Replacing~\eqref{eq:kappa1} in~\eqref{eq:DV1_1}, we obtain
\begin{xequation}
\label{eq:DV1_2}
\begin{aligned}
\grad&\V[2,\theta](x,\xc[2])\tp f_{cl,2}=\grad\V[0](x,\xc)\tp 
\bmtx{\fmap[\theta](x,\xc,u)\\ f_c}\\
&-(\theta-\th)\tp\Proj(\ff(x,\xc[2]),\th)\\
&-\ku(u-\kappa[1](x,\xc[1]))\tp\Gamma[2]\inv(u-\kappa[1](x,\xc[1]))\\
&-(u-\kappa[1](x,\xc[1]))\tp\psi[u](x,\xc)\tp\grad[x]\V[0](x,\xc)\\
&-(u-\kappa[1](x,\xc[1]))\tp\Gamma[2]\inv\D[x](\kappa[1](x,\xc[1]))\psi[theta](x,\xc)(\theta-\th).
\end{aligned}
\end{xequation}
It follows from~\ref{ass:tildebnd} and~\eqref{eq:DV1_2} that
\begin{xequation}
\label{eq:DV1_3}
\begin{aligned}
\grad&\V[2,\theta](x,\xc[2])\tp f_{cl,2}\leq\grad\V[0](x,\xc)\tp 
\bmtx{\fmap[\theta](x,\xc,u)\\ f_c}\\
&-(\theta-\th)\tp \ff(x,\xc[2])\\
&-\ku(u-\kappa[1](x,\xc[1]))\tp\Gamma[2]\inv(u-\kappa[1](x,\xc[1]))\\
&-(u-\kappa[1](x,\xc[1]))\tp\psi[u](x,\xc)\tp\grad[x]\V[0](x,\xc)\\
&-(u-\kappa[1](x,\xc[1]))\tp\Gamma[2]\inv\D[x](\kappa[1](x,\xc[1]))\psi[theta](x,\xc)(\theta-\th).
\end{aligned}
\end{xequation}
Replacing~\eqref{eq:ff} in~\eqref{eq:DV1_3}, we obtain
\begin{xequation}
\label{eq:DV1_4}
\begin{aligned}
&\grad\V[2,\theta](x,\xc[2])\tp f_{cl,2}\leq\grad\V[0](x,\xc)\tp 
\bmtx{\fmap[\theta](x,\xc,u)\\ f_c}\\
&-(\theta-\th)\tp \psi[theta](x,\xc)\tp\grad[x]\V[0](x,\xc)\\
&-\ku(u-\kappa[1](x,\xc[1]))\tp\Gamma[2]\inv(u-\kappa[1](x,\xc[1]))\\
&-(u-\kappa[1](x,\xc[1]))\tp\psi[u](x,\xc)\tp\grad[x]\V[0](x,\xc).
\end{aligned}
\end{xequation}
The control affine structure of~\eqref{eq:sysz} allows us to derive the following inequality from~\eqref{eq:DV1_4}:
\begin{xequation}
\label{eq:DV1_5}
\begin{aligned}
\grad\V[2,\theta]&(x,\xc[2])\tp f_{cl,2}\\
&\leq\grad\V[0](x,\xc)\tp 
\bmtx{\fmap[\theta](x,\xc,\kappa[1](x,\xc[1]))\\ f_c}\\
&-(\theta-\th)\tp \psi[theta](x,\xc)\tp\grad[x]\V[0](x,\xc)\\
&-\ku(u-\kappa[1](x,\xc[1]))\tp\Gamma[2]\inv(u-\kappa[1](x,\xc[1])).
\end{aligned}
\end{xequation}

%
%
%

Note that it was proved in Proposition~\ref{pro:uc0} that
\begin{xequation}\hspace{-12pt}
\begin{multlined}
\grad\V[0](x,\xc)\tp (\fmap[\theta](x,\xc,\kappa[1](x,\xc[1]))-\psi[theta](x,\xc)(\theta-\th))\\
\leq \grad\V[0](x,\xc)\tp\fmap[0](x,\xc,\kappa[0](x,\xc)),
\end{multlined}
\end{xequation}
thus, from the assumption that~$(\kappa[0],\V[0],\jset[c],\fmap[c])$ is synergistic relative to $\Amc$ for~\eqref{eq:unperturbed}, we have that}{From the assumption that~$(\kappa[0],\V[0],\jset[c],\fmap[c])$ is synergistic relative to $\Amc$ for~\eqref{eq:unperturbed}, we show in~\cite{casau_arxiv_2022} that}
\begin{multline}\label{eq:DV1_6}
\grad\V[2,\theta](x,\xc[2])\tp f_{cl,2}\leq \grad\V[0](x,\xc)\tp\fmap[0](x,\xc,\kappa[0](x,\xc))\\
-\ku(u-\kappa[1](x,\xc[1]))\tp\Gamma[2]\inv(u-\kappa[1](x,\xc[1]))\leq 0
\end{multline}
for each $(x,\xc[2])\in\Xmc\x\Xmc[c,2]$ satisfying $\V[2,\theta](x,\xc[2])<+\infty$ and each $f_{cl,2}\in\fmap[cl,2](x,\xc[2])$, hence property~\ref{ass:uc} is satisfied. 
Let $\Psi[2,\theta]$ denote the largest weakly invariant subset of
\begin{align}
(\dot x,\dot{xc2})&\in\fmap[cl,2](x,\xc[2]) & (x,\xc)\in\Emc[2]
\end{align}
with 
$\Emc[2]\ceq\{(x,\xc[2])\in\Xmc\x\Xmc[c,2]:\grad\V[2,\theta](x,\xc[2])\tp f_{cl,2}=0\\
\text{ for some }f_{cl,2}\in\fmap[cl,2](x,\xc[2])\}.$
To verify that~$(\kappa[2],\V[2,\theta],\jset[c,2],\fmap[c,2])$ is synergistic relative to $\Amc[2,\theta]$ for~\eqref{eq:systt1}, we need to check that 
$\deltabnd[2]\ceq\inf\{\gap{\V[2,\theta]}(x,\xc[2]):(x,\xc[2])\in\Psi[2,\theta]\minus\Amc[2,\theta]\}>0.$
It follows from~\eqref{eq:DV1_6} that 
$\Psi[2,\theta]\subset\{(x,\xc[2])\in\Xmc\x\Xmc[c,2]:(x,\xc[1])\in\Psi[1,\theta],u=\kappa[1](x,\xc[1])\}$
where $\Psi[1,\theta]$ is defined in Proposition~\ref{pro:synergy0}. It follows from~\eqref{eq:Amc1} that 
\begin{equation}
\begin{aligned}
&\begin{multlined}\deltabnd[2]\geq\inf\{\gap{\V[2,\theta]}(x,\xc[2]):(x,\xc[1])\in\Psi[1,\theta]\minus\Amc[1,\theta],\\ u=\kappa[1](x,\xc[1])\}\end{multlined}\\
&\phantom{\deltabnd[2]}=\inf\{\gap{\V[1,\theta]}(x,\xc[1]):(x,\xc[1])\in\Psi[1,\theta]\minus\Amc[1,\theta]\}
\end{aligned}
\end{equation}
which we have shown in Proposition~\ref{pro:synergy0} to satisfy $\deltabnd[2]>0$, under assumption~\eqref{eq:level}. In addition, $\gap{\V[2,\theta]}(x,\xc[2])=\gap{\V[1,\theta]}(x,\xc[1])\geq\gap{\V[0]}(x,\xc)>\delta(x,\xc)$ for each $(x,\xc[2])\in\Psi[2,\theta]\minus\Amc[2,\theta]$, hence the hybrid controller~$(\kappa[2],\V[2,\theta],\jset[c,2],\fmap[c,2])$ is synergistic relative to $\Amc[2,\theta]$ for~\eqref{eq:sysz} with synergy gap exceeding $\delta$.
\end{proof}

To finalize the design of a robust synergistic controller, we provide the construction of the jump map $\jmap[c,2]$ in the next proposition.

\begin{proposition}
Suppose that the sets $\Xmc$, $\Xmc[c]$, $\Umc$, and the set-valued map $\fmap[\theta]$ in~\eqref{eq:sysz} satisfy Assumption~\ref{ass:data}, and that Assumption~\ref{ass:affine} holds. Given $\Omega$ in~\eqref{eq:Omega}, a compact set $\Amc\subset\Xmc\x\Xmc[c]$, and a hybrid controller~$(\kappa[0],\V[0],\jset[c],\fmap[c])$ that is nominally synergistic relative to $\Amc$ for~\eqref{eq:sysz} with synergy gap exceeding $\delta$, $\Ascr[2]\ceq\{\Amc[2,\theta]\}_{\theta\in\Omega}$ with $\Amc[2,\theta]$ in~\eqref{eq:Amc1}, $\Vscr[2]\ceq\{\V[2,\theta]\}_{\theta\in\Omega}$ with $\V[2,\theta]$ in~\eqref{eq:V1}, the hybrid controller~$(\kappa[2],\Vscr[2],\jset[c,2],\fmap[c,2], \jmap[c,2])$ where 
\begin{multline}
\label{eq:jmapc1}
\jmap[c,2](x,\xc[2])\ceq\{(g_{c,1},g_u)\in\jset[c,2](x, \xc[2]):\\
g_{c,1}\in\jmap[c,1](x,\xc[1])\}
\end{multline}
for each $(x,\xc[2])\in\Xmc\x\Xmc[c,2]$ is synergistic relative to $\Ascr[2]$ for~\eqref{eq:sysz} with robustness margin $\Omega$ and synergy gap exceeding $\delta$.
\end{proposition}
\begin{proof}
In Proposition~\ref{pro:synergy1} we demonstrate that the hybrid controller~$(\kappa[2],\V[2,\theta],\jset[c,2],\fmap[c,2])$ is synergistic relative to $\Amc[2,\theta]$ with synergy gap exceeding $\delta$ as required by Definition~\ref{def:robust}. The proof that~\ref{ass:Vtheta} is satisfied follows closely the proof of Proposition~\ref{pro:robust0}, hence it is omitted here. The outer semicontinuity and local boundedness of $\jmap[c,2]$ follows from outer semicontinuity and local boundedness of $\jmap[c,1]$ in addition to the continuity of $\kappa[1]$, thus~\ref{ass:Gc_regular} is verified.
For each $(x,\xc[2])\in\Xmc\x\Xmc[c,2]$ and for each $g_{c,2}\in\Xmc[2]$, we have that
\begin{equation}\label{eq:gap2}
\begin{multlined}
\V[2,\theta](x,\xc[2])-\V[2,\theta](x,g_{c,2})\\
\geq \min_{\theta\in\Omega} \V[2,\theta](x,\xc[2])-\V[2,\theta](x,g_{c,2}).
\end{multlined}
\end{equation}
From~\eqref{eq:jmapc1}, it follows that $g_{c,2}\ceq(g_{c,1},g_u)$ with $g_{c,1}$ belonging to~\eqref{eq:jmapc0} and $g_u=\kappa[1](x,g_{c,1})$. Replacing~\eqref{eq:V1} in~\eqref{eq:gap2} and plugging in the aforementioned values of $g_{c,1}$ and $g_u$, we have that
\begin{equation}
\begin{aligned}\label{eq:gap2_2}
    \V[2,\theta](x,\xc[2])&-\V[2,\theta](x,g_{c,2})\\
    \geq& \max_{g_{c,1}\in\jset[c,1](x,\xc[1])}\min_{\theta\in\Omega}\V[1,\theta](x,\xc[1])-\V[1,\theta](x,g_{c,1})\\
    &+\frac{1}{2}(u-\kappa[1](x,\xc[1]))\tp\Gamma[2]\inv(u-\kappa[1](x,\xc[1]))\\
    =&\max_{g_{c,2}\in\jset[c,2](x,\xc[2])}\min_{\theta\in\Omega} \V[2,\theta](x,\xc[2])-\V[2,\theta](x,g_{c,2})
\end{aligned}
\end{equation}
for each $(x,\xc[2])\in\Xmc\x\Xmc[c,2]$ and each $g_{c,2}\ceq (g_{c,1},g_u)\in\jmap[c,2](x,\xc[2])$. Since the $\max$ and $\min$ operators in~\eqref{eq:gap2_2} commute as shown in the proof of Proposition~\ref{pro:robust0}, it follows that
\begin{align*}
\V[2,\theta](x,\xc[2])&-\V[2,\theta](x,g_{c,2})\geq \min_{\theta\in\Omega}\gap{\V[2,\theta]}(x,\xc[2])
\end{align*}
for each $(x,\xc[2])\in\Xmc\x\Xmc[c,2]$ and each $g_{c,2}\ceq (g_{c,1},g_u)\in\jmap[c,2](x,\xc[2])$, thus verifying~\ref{ass:Gc_decreases}.
\end{proof}

The hybrid closed-loop system resulting from the interconnection between~$(\kappa[2],\Vscr[2],\jset[c,2],\fmap[c,2],\jmap[c,2])$ and~\eqref{eq:sysz} is given by:
\begin{subequations}
\mathtoolsset{showonlyrefs=false}
\label{eq:sys1Omega}
\begin{align}
(\dot x,\dot{xc2})&\in\fmap[cl,2](x,\xc[2]) & (x,\xc[2])&\in\fset[\Omega,2]\\
(x\pl, \xc[2]\pl)&\in\jmap[\Omega,2](x,\xc[2]) & (x,\xc[2])&\in\jset[\Omega,2]
\end{align}
\end{subequations}
where\par
\resizebox{0.48\textwidth}{!}{\hspace*{-10.2pt}\parbox{0.5\textwidth}{\begin{align*}
\fset[\Omega,2]&\ceq\left\{(x,\xc[2])\in\Xmc\x\Xmc[c,2]:\min_{\theta\in\Omega}\gap{\V[2,\theta]}(x,\xc[2])\leq\delta(x,\xc)\right\}\\
\jset[\Omega,2]&\ceq\left\{(x,\xc[2])\in\Xmc\x\Xmc[c,2]:\min_{\theta\in\Omega}\gap{\V[2,\theta]}(x,\xc[2])\geq\delta(x,\xc)\right\}
\end{align*}}}
and
\begin{align*}
\jmap[\Omega, 2](x,\xc[2])\ceq\bmtx{x\\ \jmap[c,2](x,\xc[2])} \hfill \forall (x,\xc[2])\in\jset[\Omega,2].
\end{align*}

The global asymptotic stability of $\Amc[2,\theta]$ for~\eqref{eq:sys1Omega} follows from Theorem~\ref{thm:rgas} and it is stated in the next corollary for the sake of completeness. The proof is omitted because it is identical to the proof of Corollary~\ref{cor:gas0}

\begin{corollary}
Suppose that the sets $\Xmc$, $\Xmc[c]$, $\Umc$, and the set-valued map $\fmap[\theta]$ in~\eqref{eq:sysz} satisfy Assumption~\ref{ass:data}, and that Assumption~\ref{ass:affine} holds. Given $\Omega$ in~\eqref{eq:Omega}, a positive function $\delta:\Xmc\x\Xmc[c]\mapsto\R{}$, a compact set $\Amc\subset\Xmc\x\Xmc[c]$, and a hybrid controller~$(\kappa[0],\V[0],\jset[c],\fmap[c])$ that is nominally synergistic relative to $\Amc$ for~\eqref{eq:sysz} with synergy gap exceeding $\delta$, for each $\theta\in\Omega$, the set $\Amc[2,\theta]$ is globally asymptotically stable for~\eqref{eq:sys1Omega}.
\end{corollary}

In the next section, we apply the controllers proposed in Sections~\ref{sec:start} and~\ref{sec:back} to global asymptotic stabilization of a setpoint for a two-dimensional system in the presence of an obstacle.

\section{Synergistic Hybrid Feedback for Robust Global Obstacle Avoidance}\label{sec:obstacles}
To demonstrate the applicability of the synergistic adaptive controller of Section~\ref{sec:adaptive}, we consider the problem of globally asymptotically stabilizing the origin for a vehicle moving on a plane with an obstacle
$\Nmc\ceq z_0+\radius\cl{\ball}$
with $z_0\in\R 2$ and $\radius>0$ such that the origin is not contained in $\Nmc$.
We consider that the evolution in time of the position $z\in\R 2\minus\Nmc$ of the vehicle is described by 
\begin{equation}
    \label{eq:zobs}
    \dot z=u+\theta
\end{equation}
where $u\in\R{2}$ is the input and $\theta\in\R 2$ is an unknown constant. 
We have shown in~\cite[Section~IV]{Casau2019} that
$\psi(z)\ceq\bmtx{\log(\norm{z-z_0}-\radius)\ \frac{z-z_0}{\norm{z-z_0}}}\tp$
is a diffeomorphism between $\R 2\minus\Nmc$ and $\R{}\x\sphere{1}$, hence global asymptotic stabilization of the origin for~\eqref{eq:zobs} is equivalent to the global asymptotic stabilization of $\psi(0)$. 
for 
\begin{equation}\label{eq:xobs}
    \dot x=\D \psi(\psi\inv(x))u+\D \psi(\psi\inv(x))\theta.    
\end{equation}
with $x\in\Xmc\ceq\R{}\x\sphere{1}$. Before moving to the controller design, we show that Assumption~\ref{ass:data} is verified for the particular problem at hand.
\begin{proposition}
The sets $\Xmc\ceq\R{}\x\sphere{1}$, $\Xmc[c]\ceq\{-1,1\}$, $\Umc\ceq\R{2}$ and the set-valued map 
\begin{equation}
    \label{eq:ftt_sim}
    \fmap[\theta](x,\xc,u)\ceq\D \psi(\psi\inv(x))u+\D \psi(\psi\inv(x))\theta
\end{equation}
defined for each $(x,\xc,u)\in\Xmc\x\Xmc[c]\x\Umc$ satisfies Assumption~\ref{ass:data} and
\begin{enumerate}[label=($\star$)]
    \item\label{star}The intersection between $\fmap[\theta](x,\xc,u)$ and the tangent space to $\Xmc$ at $(x,\xc,u)$ is nonempty for each $(x,\xc,u)\in\Xmc\x\Xmc[c]\x\Umc$.
\end{enumerate}
\end{proposition}
\begin{proof}
    To check that the condition~\ref{ass:star} holds, note that the sets $\Xmc$, $\Xmc[c]$ and $\Umc$ are closed subsets of $\R{3}$, $\R{}$ and $\R{2}$, respectively.
    It follows from the fact that $\psi$ is a diffeomorphism between $\R{2}\minus\Nmc$ and $\R{}\x\sphere 1$ that $\D\psi(\psi\inv(x))$ is an isomorphism between the tangent space to $\R{2}\minus\Nmc$ at $\psi\inv(x)$ and the tangent space to $\R{}\x\sphere 1$ at $x$ for each $x\in\Xmc$ (cf.~\cite[Proposition~3.6]{lee_introduction_2000}), thus~\ref{star} is verified. 
    Since $\psi$ is a diffeomorphism it also follows that $x\mapsto\D\psi(\psi\inv(x))$ is continuous, thus $\fmap[\theta]$ is also continuous and single-valued, hence it verifies~\ref{ass:Fbasic}.
\end{proof}

\begin{remark}\label{rem:vcs}
    The condition~\ref{star} is pivotal in the verification of the conditions~\ref{ass:VC} and~\ref{ass:VC2} for this particular example, which, in turn, allows us to check the completeness of maximal solutions as shown in Theorems~\ref{thm:gas} and~\ref{thm:rgas}, respectively.
\end{remark}

The controller design of Section~\ref{sec:adaptive} requires the existence of a hybrid controller of the form~$(\kappa[0],\V[0],\jset[c],\fmap[c])$ that is nominally synergistic relative to 
\begin{equation}
    \label{eq:Amc_sim}
    \Amc\ceq\{(x,q)\in\Xmc\x\Xmc[c]:x=\psi(0)\}.
\end{equation}
for~\eqref{eq:ftt_sim}, thus we start by showing that the controller provided in~\cite[Section~IV]{Casau2019} satisfies the requirements~\ref{ass:Qbasic}-\ref{ass:Psi}.
In this direction, let the controller variable $\xc$ in~\eqref{eq:sysz} be a logic variable $q$ which is either $1$ or $-1$ and whose values does not change during flows, i.e., $\xc=q\in\Xmc[c]\ceq\{-1,1\}$ and $\dot{q}=\fmap[c](x,q)\ceq 0$ for all $(x,q)\in\Xmc\x\Xmc[c]$ which verifies~\ref{ass:Fc}.
Following the controller design of~\cite[Section~IV]{Casau2019}, let
    $\phi[q](x)\ceq\bmtx{x_1& \frac{x_2}{1-q x_3}}\tp$
for each $x\ceq(x_1,x_2,x_3)\in\U[q]\ceq\{x\in\Xmc: q x_3\neq 1\}$ with $q\in\Xmc[c]\ceq\{-1,1\}$. Furthermore, we define 
\begin{equation}
    \label{eq:Vobs}
    \V[0](x,q)\ceq\begin{dcases}
        \frac{1}{2}\norm{\phi[q](x)-\phi[q](\psi(0))}^2 & \text{ if }x\in\U[q]\\
        +\infty & \text{otherwise}
    \end{dcases}
\end{equation}
for each $(x,q)\in\Xmc\x\Xmc[c]$. Defining $\jset[c](x,q)=\Xmc[c]$ for each $(x,q)\in\Xmc\x\Xmc[c]$ and noting that $\{(\U[q],\phi[q])\}_{q\in\Xmc[c]}$ covers $\R{}\x\sphere{1}$ we have that the optimization problem in~\eqref{eq:mins} is feasible, hence assumption~\ref{ass:Qbasic} is verified.
Since each chart $\phi[q]:\U[q]\to\R 2$ is a diffeomorphism, $\phi[q](x)=\phi[q](\psi(0))$ if and only if $x=\psi(0)$, hence $\V$ in~\eqref{eq:Vobs} is positive definite relative to~\eqref{eq:Amc_sim}.
Moreover, $\V[0]$ is continuous and $\V[0]\inv([0,c])$ is compact for each $c\in\Rnneg$, thus~\ref{ass:V} is verified. Since $\jset[c]$ is constant and equal to the finite set $\Xmc[c]$ for each $(x,\xc)\in\Xmc\x\Xmc[c]$, we have that $\jset[c]$ is outer semicontinuous, lower semicontinuous and locally bounded, hence~\ref{ass:Q} is verified. Condition~\ref{ass:kappa} is verified for
\begin{equation*}
    \kappa[0](x,q) =
     -\left(\D \psi(\psi\inv(x))\right)\tp\D\phi[q](x)\tp(\phi[q](x)-\phi[q](\psi(0)))
\end{equation*}
for each $(x,q)\in\dom\kappa[0]=\{(x,q)\in\Xmc\x\Xmc[c]:x\in\U[q]\}.$ The previous arguments allow us to make the following assertion.
\begin{proposition}
Given $\Amc$ in~\eqref{eq:Amc_sim}, the hybrid controller~$(\kappa[0],\V[0],\jset[c],\fmap[c])$ is a synergistic candidate relative to $\Amc$ in~\eqref{eq:Amc_sim} for~\eqref{eq:xobs}.
\end{proposition}

Even though~\eqref{eq:Vobs} is continuous, it is not Lipschitz continuous everywhere, hence the proof that~\ref{ass:uc} holds might not be immediately obvious. From~\eqref{eq:gap} and using the fact that $\jset[c](x,q)\ceq\Xmc[c]\ceq\{-1,1\}$ for each $(x,q)\in\Xmc\x\Xmc[c]$, we have that 
$\gap{\V[0]}(x,q)=\max\{0,\V[0](x,q)-\V[0](x,-q)\}$
for each $(x,q)\in\Xmc\x\Xmc[c]$ and, in particular, we have that $\gap{\V[0]}(x,q)=+\infty$ for each $(x,q)\not\in\U[q]$, hence, for any function $\delta$, it follows that each $(x,q)\in\Xmc\x\Xmc[c]$ satisfying $(x,q)\in\U[q]$ does not belong to $\fset$. Since $\U[q]$ is open relative to $\Xmc\ceq\R{}\x\sphere{1}$ for each $q\in\Xmc[c]$, $\{(x,q)\in\Xmc\x\Xmc[c]:x\in\Xmc\minus\U[q]\}$ and $\fset$ are disjoint closed sets, and there exists a neighborhood of $\fset$ where $\V[0]$ is Lipschitz continuous. The generalized derivative of $\V[0]$ at $(x,q)$ is the direction $\fmap[cl,0](x,q)$ is given by
\begin{multline}\label{eq:clarke_Vn_obs}
\clarke{\V[0]}(x,q;\fmap[cl,0](x,q))=\\-\norm{\D\psi(\psi\inv(x))\tp\D\phi[q](x)\tp(\phi[q](x)-\phi[q](\psi(0)))}^2,
\end{multline}
for each $(x,q)\in\fset$, where $\fmap[cl,0]$ is the flow map for the closed-loop system resulting from the interconnection between~$(\kappa[0],\V[0],\jset[c],\fmap[c])$ and~\eqref{eq:xobs} with $\theta=0$ (cf.~\eqref{eq:closed}). It follows from~\eqref{eq:clarke_Vn_obs} that the growth of $\V[0]$ along the flows of the closed-loop system is upper bounded by $0$, hence~\ref{ass:uc} is verified.
It follows from the fact that $\psi$ and $\{\phi[q]\}_{q\in\Xmc[c]}$ are diffeomorphisms that Assumption~\ref{ass:Psi} is satisfied with $\Psi=\Amc$ (cf.~\cite{Casau2019}), thus the following holds. 

\begin{proposition}
    Given $\Amc$ in~\eqref{eq:Amc_sim} and a continuous function $\delta:\Xmc\x\Xmc[c]\to\R{}$, the hybrid controller~$(\kappa[0],\V[0],\jset[c],\fmap[c])$ is nominally synergistic relative to $\Amc$ in~\eqref{eq:Amc_sim} for~\eqref{eq:xobs} with synergy gap exceeding $\delta$.
\end{proposition}

An additional property of the kind of synergistic feedback presented above is that any function $\delta$ satisfies~\ref{ass:deltapsi} since $\Psi\minus\Amc=\emptyset$. Therefore, any choice of $\delta$ satisfying~\ref{ass:deltazero} yields global asymptotic stability for the hybrid closed-loop system as proved in Theorem~\ref{thm:gas}. Since Assumption~\ref{ass:affine} is satisfied, we meet all the requirements for the controller design of Section~\ref{sec:adaptive}, thus it is only a matter of applying the procedures described therein to obtain an adaptive synergistic hybrid feedback controller that is able to deal with parametric uncertainty. 
In the next section, we present some numerical results that illustrate the behaviour of the closed-loop system.

\subsection{Simulation Results}\label{sec:sims}
In this section, we present simulation results of the closed-loop system resulting from the interconnection between~\eqref{eq:xobs} and the hybrid controllers that are presented in Section~\ref{sec:adaptive} considering that there is an obstacle $\Nmc\ceq z_0+\radius\cl{\ball}$ with $z_0=\bmtx{1 & 0}\tp$ and $\radius=0.5$. Furthermore, we consider that $\theta=\bmtx{\sqrt{2}/2 & \sqrt{2}/2}\tp$ and that the controller parameters are $\ku=1$, $\Gamma[1]=\Gamma[2]=\eye{2}$, $\slack=1$, $\theta_0=1$, and $\delta(x,q)=1$ for each $(x,q)\in\Xmc\x\Xmc[c]$.
For this particular choice of $\Gamma[1]$, we have that
\IfArxiv{$$\hat{\jmap}(\th)=\begin{cases}
\th & \text{ if } \norm{\th}\leq\tz\\
\tz\frac{\th}{\norm{\th}} & \text{otherwise}
\end{cases}$$}{$\hat{\jmap}(\th)=\min\{1,\tz/\norm{\th}\}\th$}
for each $\th\in\Omega+\slack\cl{\ball}$, which is outer semicontinuous and locally bounded.

\begin{figure}
    \input{traj.tex}
    \includegraphics[width=0.49\textwidth,right]{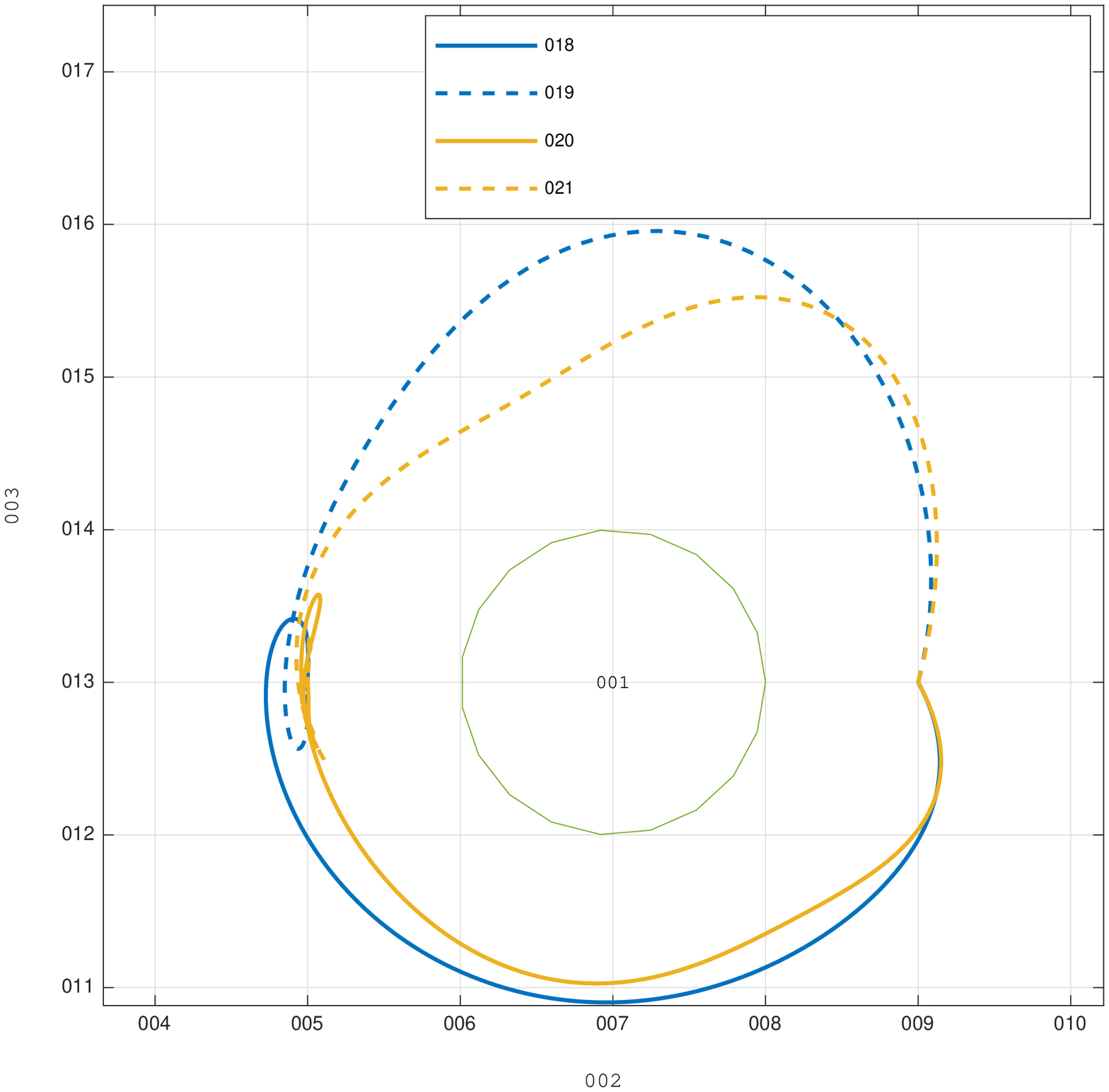}
    \caption{Trajectories $t\mapsto z(t)$ of~\eqref{eq:zobs} for the closed-loop system with parameters given in Section~\ref{sec:sims}.}
    \label{fig:traj}
\end{figure}

\begin{figure}
    \input{plots.tex}
    \includegraphics[width=0.48\textwidth,right]{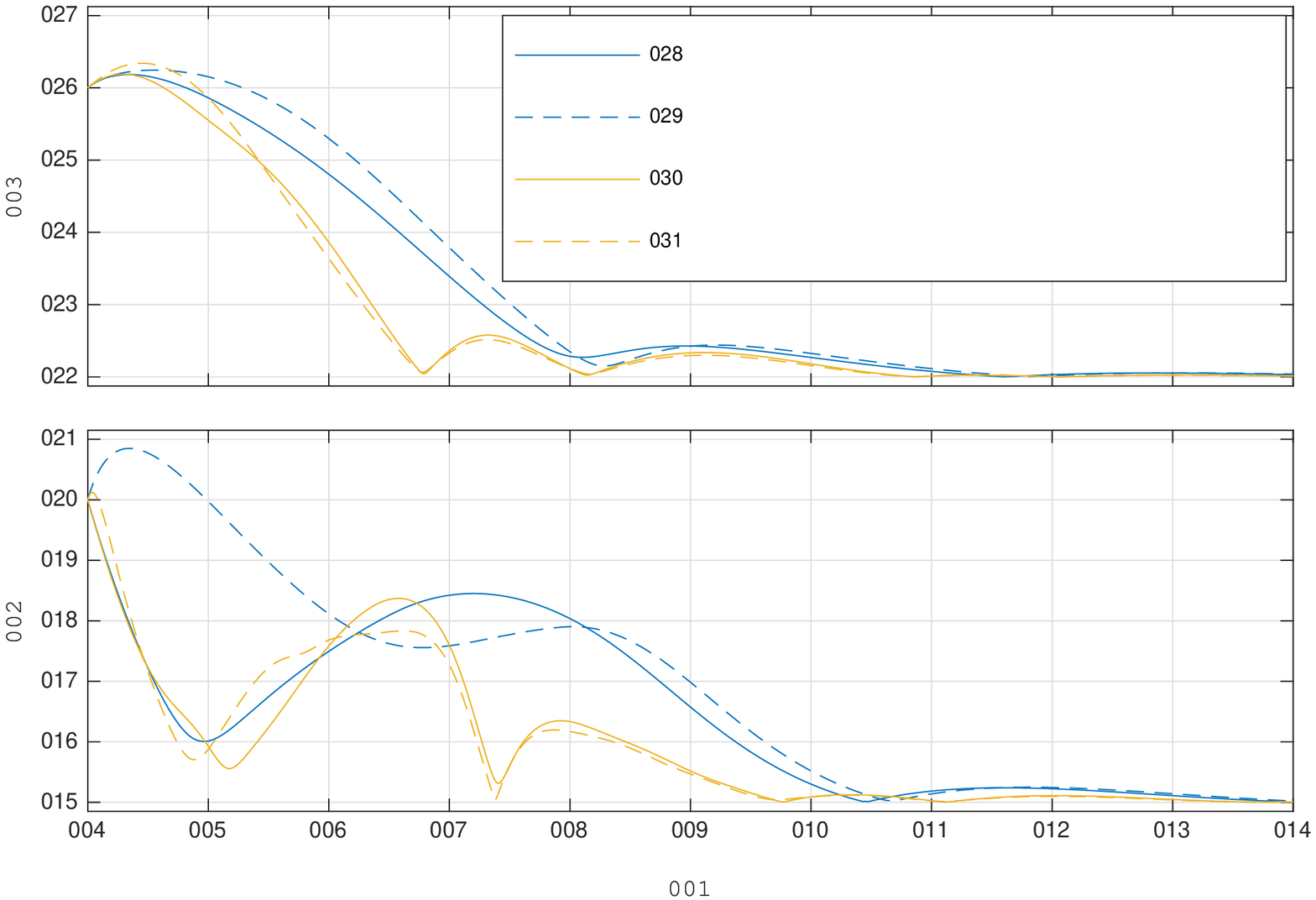}
    \caption{Evolution in time of the distance to the origin (top) and the parametric estimation error (bottom) for the closed-loop system with parameters given in Section~\ref{sec:sims}.}
    \label{fig:plots}
\end{figure}

Figure~\ref{fig:traj} represents the trajectory of the vehicle starting from rest at $z(0) = \bmtx{2 & 0}\tp$ for each of the controllers presented in Section~\ref{sec:adaptive}. It can be verified both through Figure~\ref{fig:traj} as well as Figure~\ref{fig:plots} that the trajectories before and after backstepping are comparable, since the evolution of the distance of the vehicle to the desired setpoint is fairly similar in both cases. The bottom half of Figure~\ref{fig:plots} depicts the evolution of the estimation error, which has a smaller settling time for the closed-loop system with the controller of Section~\ref{sec:back} than the controller of Section~\ref{sec:start} for this particular simulation. To find out more about the simulation and its implementation, you may explore the source code at \url{https://github.com/pcasau/synergistic}.

\section{Conclusion}\label{sec:conclusion}
Synergistic hybrid feedback has taken many forms over the years, depending on the particular dynamical system being studied. The unifying framework for synergistic hybrid feedback that we presented in this paper captures the most salient features of existing synergistic hybrid feedbacks in order to help others distinguish between the particular and the general in different instances of synergistic hybrid feedback across the literature. In addition, we provided a controller design that starts from an existing synergistic controller and modified it in order to yield an adaptive controller that is able to compensate for the presence of bounded matched uncertainties in affine control systems. Furthermore, we demonstrated that the proposed controller is amenable to backstepping and can be applied to the problem of global obstacle avoidance.

\bibliographystyle{ieeetr}
\bibliography{biblio}

\begin{IEEEbiography}[{\includegraphics[width=1in,height=1.25in,clip,keepaspectratio]{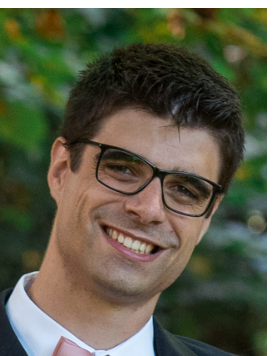}}]{Pedro Casau}
    is a Research Fellow at the Institute for Systems and Robotics, Lisbon, Portugal. He received the B.Sc. in Aerospace Engineering in 2008 from Instituto Superior Técnico (IST), Lisbon, Portugal.  In 2010, he received the M.Sc. in Aerospace Engineering from IST and enrolled in the Electrical and Computer Engineering Ph.D. program at the same institution which he completed  with distinction and honours in 2016. He participated on several national and international research projects on guidance, navigation and control of unmanned air vehicles (UAVs) and satellites. His current research interests include nonlinear control, hybrid control systems, vision-based control systems, controller design for autonomous air-vehicles. 
    \end{IEEEbiography}\vspace*{-12pt}
    \begin{IEEEbiography}[{\includegraphics[width=1in,height=1.25in,clip,keepaspectratio]{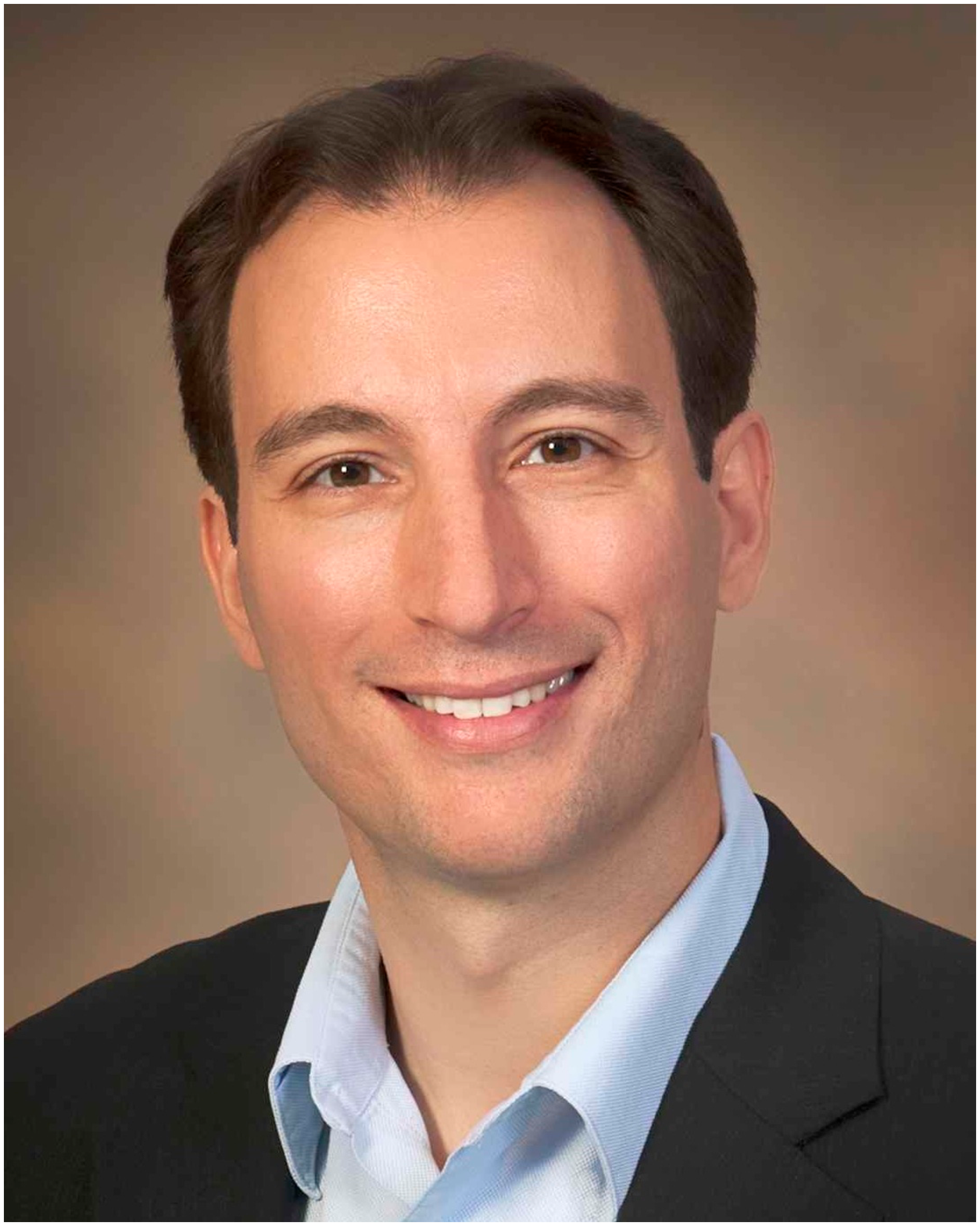}}]{Ricardo G. Sanfelice} received the B.S. degree in Electronics Engineering from the Universidad de Mar del Plata, Buenos Aires, Argentina, in 2001, and the M.S. and Ph.D. degrees in Electrical and Computer Engineering from the University of California, Santa Barbara, CA, USA, in 2004 and 2007, respectively. In 2007 and 2008, he held postdoctoral positions at the Laboratory for Information and Decision Systems at the Massachusetts Institute of Technology and at the Centre Automatique et Systèmes at the École de Mines de Paris. In 2009, he joined the faculty of the Department of Aerospace and Mechanical Engineering at the University of Arizona, Tucson, AZ, USA, where he was an Assistant Professor. In 2014, he joined the University of California, Santa Cruz, CA, USA, where he is currently Professor in the Department of Electrical and Computer Engineering. Prof. Sanfelice is the recipient of the 2013 SIAM Control and Systems Theory Prize, the National Science Foundation CAREER award, the Air Force Young Investigator Research Award, the 2010 IEEE Control Systems Magazine Outstanding Paper Award, and the 2020 Test-of-Time Award from the Hybrid Systems: Computation and Control Conference. He is Associate Editor for Automatica and a Fellow of the IEEE. His research interests are in modeling, stability, robust control, observer design, and simulation of nonlinear and hybrid systems with applications to power systems, aerospace, and biology.
    \end{IEEEbiography}\vspace*{-12pt}
    \begin{IEEEbiography}[{\includegraphics[width=1in,height=1.25in,clip,keepaspectratio]{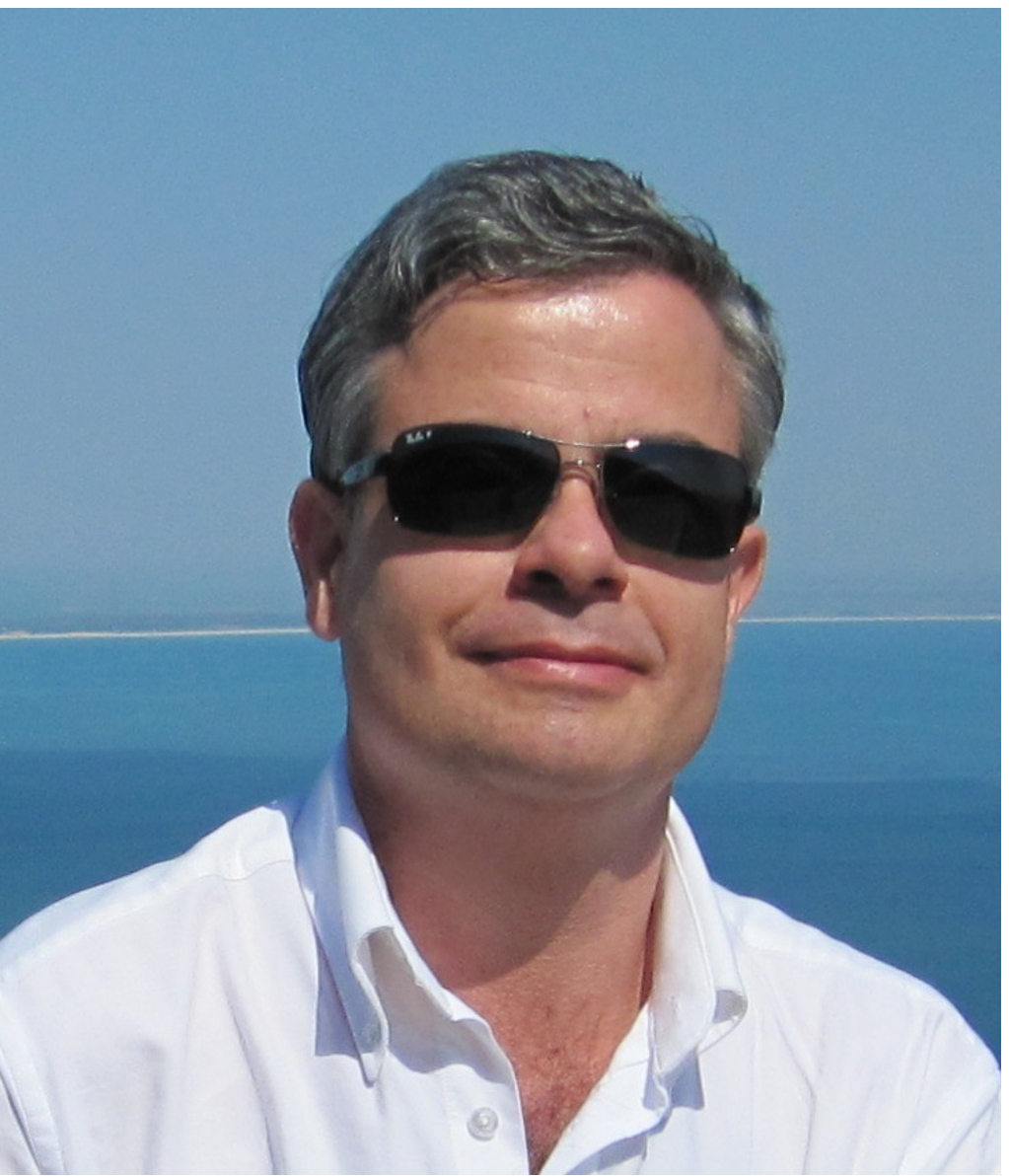}}]{Carlos Silvestre}
     received the Licenciatura degree in Electrical Engineering from the Instituto Superior Tecnico (IST) of Lisbon, Portugal, in 1987 and the M.Sc. degree in Electrical Engineering and the Ph.D. degree in Control Science from the same school in 1991 and 2000, respectively. In 2011 he received the Habilitation in Electrical Engineering and Computers also from IST. Since 2000, he is with the Department of Electrical Engineering of the Instituto Superior Tecnico, where he is currently an Associate Professor of Control and Robotics on leave. Since 2015 he is a Professor of the Department of Electrical and Computers Engineering of the Faculty of Science and Technology of the University of Macau. Over the past years, he has conducted research on the subjects of navigation, guidance and control of air and ocean robots. His research interests include linear and nonlinear control theory, hybrid control, sensor based control, coordinated control of multiple vehicles, networked control systems, fault detection and isolation, and fault tolerant control.
    \end{IEEEbiography}

\end{document}

%% file: packages.tex
\usepackage[utf8]{inputenc}
\usepackage{amsmath}
\usepackage{amsfonts}
\usepackage{amssymb}
\usepackage{mathrsfs}
\usepackage{paralist}
\usepackage{psfrag}
\usepackage{pstool}

\usepackage{enumitem}
\usepackage{mathtools}

\usepackage{amsthm}
\usepackage{xcolor}
\usepackage{hyperref}
\usepackage{comment}
\usepackage{xstring}
\usepackage{mdwlist}
\usepackage[export]{adjustbox}

%% file: macros.tex
\newcommand{\pedro}[1]{{\color{red} #1}}
\renewcommand{\tilde}[1]{\widetilde{#1}}
\newcommand{\lbar}[1]{\underline{#1}}

\newcommand{\ceq}{:=}
\newtheorem{theorem}{Theorem}
\newtheorem{proposition}{Proposition}
\newtheorem{corollary}{Corollary}
\newtheorem{lemma}{Lemma}
\theoremstyle{definition}
\newtheorem{assumption}{Assumption}

\newtheorem{definition}{Definition}

\newtheorem{remark}{Remark}
\newcommand{\pl}{^+}
\DeclareMathOperator*{\argmin}{\arg\,\min}
\DeclareMathOperator*{\argmax}{\arg\,\max}
\newcommand{\pmtx}[1]{\begin{pmatrix}#1\end{pmatrix}}
\newcommand{\bmtx}[1]{\begin{bmatrix}#1\end{bmatrix}}
\newcommand{\pder}[2]{\frac{\partial #1}{\partial #2}}
\newcommand{\x}{\times}
\newcommand{\inv}{^{-1}}
\newcommand{\minus}{\backslash}

\newcommand{\R}[1]{\mathbb{R}^{#1}}
\newcommand{\Rnneg}{\R{}_{\geq 0}}
\DeclareMathOperator{\bd}{bd}
\newcommand{\tp}{^\top}
\newcommand{\inner}[2]{\langle #1,#2\rangle}
\newcommand{\norm}[2][\mbox{}]{\left| #2\right|_{#1}}

\newcommand{\eye}[1]{I_{#1}}
\let\vec\relax
\DeclareMathOperator{\vec}{vec}

\DeclareMathOperator{\dom}{dom}

\newcommand{\grad}[1][]{\nabla_{#1}}

\newcommand{\so}[1]{\mathfrak{so}(#1)}

\newcommand{\sphere}[1]{\mathsf{S}^{#1}}

\newcommand{\kron}[2]{#1 \otimes #2}

\newcommand{\D}[1][\mbox{}]{\mathcal{D}_{#1}}

\newcommand{\projt}{{\downarrow_{t}}}

\newcommand{\Amc}{\mathcal{A}}
\newcommand{\ball}{\mathbb{B}}

\DeclareMathOperator{\cl}{cl}

\DeclareMathOperator{\rge}{rge}

%% file: notation.tex
\IfArxiv{\subsection{Topology, Metric Spaces, Functions, and Set-Valued Maps}}{}

\IfArxiv{\hypertarget{def:hausdorff}{G}iven a topological space $X$, a neighborhood of a set $S$ is any open set that contains $S$. A topological space $X$ is said to be Hausdorff if, given any pair of distinct points $q_1,q_2\in X$, there exist neighborhoods $U_1$ of $q_1$ and $U_2$ of $q_2$ that do not intersect. Any metric space is Hausdorff, hence the Euclidean spaces are Hausdorff. Lemma~4.29 in~\cite{lee_introduction_2000} points out that any closed subspace of an locally compact Hausdorff space is itself locally compact Hausdorff. A set is said to be locally compact if for each point there is a neighborhood which is precompact, i.e., whose closure is a compact set.

\hypertarget{def:subspace}{A} topology on a set $X$ is a collection $T$ of subsets of $X$, called open sets, satisfying the following properties: $X$ and $\emptyset$ are elements of $T$; $T$ is closed under finite intersections; and $T$ is closed under arbitrary unions.
The subspace topology of a subset $A$ of $X$ is the collection of subsets of $A$ that are obtained from the intersection of $A$ with an open set of $X$. A subset $A$ of a topological space $X$ that is endowed with the subspace topology is said to be a subspace of $X$.

\hypertarget{def:metric}{A} metric space is a set $M$ together with a metric $d$. A set $S\subset M$ is open in the metric space sense if for each $x\in S$ there exists $\epsilon>0$ such that the set points $y\in M$ satisfying $d(x,y)<\epsilon$ are contained in $S$. The metric topology on $M$ is the collection of all subsets of $M$ that are open in the metric space sense (cf.~\cite[Exercise~2.1]{lee_introduction_2000}).}{}

\hypertarget{def:Euclid}{T}he Cartesian Product $\R{n}=\R{}\x\ldots\R{}$ of $n$ copies of the real line together with scalar multiplication and component-wise addition of vectors is known as $n$-dimensional Euclidean space. The Euclidean metric topology is the one induced by the metric $x\mapsto\norm{x}\ceq\sqrt{x\tp x}$. The $n$-dimensional Euclidean space has the topology generated by a countable basis of open balls of the form $c+\slack\ball\ceq\{x\in\R n:\norm{x-c}<\slack\}$, where $c\in\R n$ and $\slack>0$. More generally, given a set $\Omega\subset\R n$, we define $\Omega+\slack\ball\ceq\bigcup_{c\in\Omega}c+\slack\ball$. The operators $\bd S$ and $\cl{S}$ denote the boundary and the closure of a set $S$, respectively.

Given a function $f:\R m\to\R n$, the preimage of a set $U\subset\R n$ through $f$ is 
$f\inv(U)\ceq\{x\in\R m: f(x)\in U\}.$
Similarly, the image of a set $W$ through $f$ is 
$f(W)\ceq\{y\in\R n:y=f(x)\text{ for some }x\in W\}.$

\hypertarget{def:setvaluedmap}{A} set-valued map $M$ from $S\subset\R{m}$ to the power set of some Euclidean space $\R{n}$ is represented by $M:S\tto\R{n}$. 
The domain of a set-valued map is given by
$\dom M\ceq\{x\in\R{n}: M(x)\neq \emptyset\}.$
Given a subset $S$ of $\R{m}$, a set-valued map $M:S\tto\R{n}$ is said to be outer semicontinuous (osc) relative to $S$ if its graph, given by 
$\gph M\ceq\{(x,y)\in S\x\R{n}:y\in M(x)\},$
is closed relative to $S\x\R{n}$. The set-valued map $M$ is locally bounded at $x\in \R m$ if there exists a neighborhood $U_x$ of $x$ such that $M(U_x)\subset\R{n}$ is bounded. It is locally bounded relative to $S$ if the set-valued mapping from $\R m$ to $\R n$ defined by $M(x)$ for $x\in S$ and $\emptyset$ for $x\not\in S$ is locally bounded at each $x\in S$. It is convex-valued if $M(x)$ is convex for each $x\in S$.

\hypertarget{def:continuous}{A} set-valued map $M:S\tto\R{n}$ is upper semicontinuous (usc) at $x$ if, for each open set $V\subset\R{n}$ that contains $M(x)$, there exists a neighborhood $U$ of $x$ such that $x'\in U\cap S$ implies $M(x')\subset V$. The map $M$ is lower semicontinuous (lsc) at $x$ if, for each open set $V\subset\R{n}$ satisfying $M(x)\cap V\neq\emptyset$, there exists a neighborhood $U$ of $x$ such that $x'\in U\cap S$ implies $M(x')\cap V\neq\emptyset$. The map $M$ is continuous at $x$ if it is both lsc and usc at $x$. The map $M$ is usc, lsc, continuous on $S$ if it is usc, lsc, continuous, respectively, at each $x\in S$.

\IfArxiv{\subsection{Differentiability}}{}

\hypertarget{def:tangentCone}{T}he tangent cone to a set $S\subset\R{n}$ at a point $x\in\R{n}$, denoted by $\T[x]S$, is the set of all vectors $w\in\R{n}$ for which there exists $x_i\in S$, $\tau_i>0$ with $x_i\to x$, $\tau_i$ convergent to $0$ from above, and $w=\lim_{i\to\infty}\frac{x_i-x}{\tau_i}.$

Given a differentiable function $F:\R{m\x n}\to\R{p\x q}$, we define $\D F(X)\ceq\pder{\vec(F)}{\vec(X)\tp}(X)$ for each $X\in\R{m\x n}$, where $\vec(A)\ceq[\ee[1]\tp A\tp\ \ldots\ \ee[m]\tp A\tp]\tp$ for each $A\in\R{m\x n}$ and $\ee[i]\in\R m$ is a vector of zeros, except for the $i$-th component, which is $1$. If $F$ has multiple arguments, say $(X,Y)\in\R{m\x n}\x\R{k\x\ell}$, we define $\D[X]F(X,Y)\ceq\pder{\vec(F)}{\vec(X)\tp}(X,Y)$ for each $(X,Y)\in\R{m\x n}\x\R{k\x\ell}$. If $F:\R{n}\to\R{}$, then $\grad F(x)\ceq\D F(x)\tp$ for each $x\in\R n$. If $F:\R{n}\x\R m\to\R{}$, then $\grad[x] F(x,y)\ceq\D[x] F(x,y)\tp$ for each $(x,y)\in\R n\x\R m$ and $\grad[y] F(x,y)\ceq\D[y] F(x,y)\tp$ for each $(x,y)\in\R n\x\R m$.

Clarke's generalized directional derivative of a function $\V:\R{n}\to\R{}$ in the direction $v$, is defined as follows (c.f.~\cite[Eq.~(1)]{Clarke1987}):
$\clarke{\V}(x;v)\ceq\limsup_{\substack{y\to x\\ \lambda\searrow 0}} \frac{\V(y+\lambda v)-\V(y)}{\lambda}.$

\IfArxiv{\subsection{Stability of Hybrid Systems}}{}

A hybrid system $\Hmc$ with state space $\R n$ is defined in~\cite{goebel_hybrid_2012} and~\cite{Sanfelice2021} as
\begin{equation}\label{eq:hybridsystem}
\begin{aligned}
\dot{\xi}&\in \fmap(\xi) & \xi&\in \fset\\
\xi\pl&\in \jmap(\xi) & \xi&\in \jset
\end{aligned}
\end{equation}
where $\xi\in\R n$ is the state, $\fset\subset\R n$ is the flow set, $\fmap:\R n\rightrightarrows\R n$ is the flow map, $\jset\subset\R n$ denotes the jump set, and $\jmap:\R n\rightrightarrows\R n$ denotes the jump map. A solution $\xi$ to $\Hmc$ is parametrized by $(t,j)$, where $t$ denotes ordinary time and $j$ denotes the jump time, and its domain $\dom \xi \subset \mathbb{R}_{\geq 0}\times \mathbb{N}$ is a hybrid time domain:  for each $(T,J)\in \dom \xi$, $\dom \xi\cap ([0,T]\times\{0,1,\dots J\})$ can be written in the form $\cup_{j=0}^{J-1}([t_j,t_{j+1}],j)$ for some finite sequence of times $0=t_0\leq t_1\leq t_2\leq \cdots \leq t_J$, where $I_j := [t_j,t_{j+1}]$ and the $t_j$'s define the jump times. \hypertarget{def:maximal}{A} solution $\xi$ to a hybrid system is said to be \emph{maximal} if it cannot be extended by flowing nor jumping and \emph{complete} if its domain is unbounded.

\hypertarget{def:invariant}{A} set $S$ is said to be forward pre-invariant for a hybrid system~\eqref{eq:hybridsystem} if each maximal solution of~\eqref{eq:hybridsystem} starting in $S$ remains in $S$. It is said to be forward invariant if it is forward pre-invariant and each maximal solution from $S$ is complete (see e.g.~\cite[Chapters~3~and~7]{Sanfelice2021}).

\hypertarget{def:hbc}{T}he hybrid basic conditions provide a set of sufficient conditions for well-posedness and they are as follows (cf.~\cite[Assumption~6.5]{goebel_hybrid_2012}):
\begin{enumerate}[label=(A\arabic*)]
    \item\label{ass:basicCD} $\fset$ and $\jset$ are closed subsets of $\R{n}$;
    \item\label{ass:basicF} $\fmap:\R n\tto\R n$ is osc and locally bounded relative to $\fset$, $\fset\subset\dom\fmap$, and $\fmap(x)$ is convex for every $x\in\fset$;
    \item\label{ass:basicG} $\jmap:\R n\tto \R n$ is osc and locally bounded relative to $\jset$, and $\jset\subset\dom\jmap$.
\end{enumerate}

Given a function $\V:\R n\to\Rnneg$ that is Lipschitz continuous on a neighborhood of $\fset$ in~\eqref{eq:hybridsystem} and $\uc:\R n\to\Rnneg$, we say that the growth of $\V$ along flows of~\eqref{eq:hybridsystem} is bounded by $\uc$ if the following holds:
\begin{align}\label{eq:fbound}
\clarke{\V}(\xi;f)&\leq\uc(\xi) & \forall\xi&\in\fset,\ \forall f\in\fmap(\xi)\cap\T[\xi]\fset.
\end{align}
If, for some function $\uc:\R n\to\Rnneg$,
\begin{align}\label{eq:jbound}
\V(\xi)-\V(\xi)&\leq\ud(\xi) & &\forall\xi\in\jset,\ \forall \xi\in\jmap(\xi),
\end{align}
then we say that the growth of $\V$ along jumps of~\eqref{eq:hybridsystem} is bounded by $\ud$. If both~\eqref{eq:fbound} and~\eqref{eq:jbound} hold, then we say that the growth of $\V$ along solutions to~\eqref{eq:hybridsystem} is bounded by $\uc,\ud$.

A compact set $\Amc$ is said to be \emph{stable} for~\eqref{eq:hybridsystem} if for every $\epsilon>0$ there exists $\delta>0$ such that every solution $\phi$ to~\eqref{eq:hybridsystem} with $\norm[\Amc]{\phi(0,0)}\leq \delta$ satisfies $\norm[\Amc]{\phi(t,j)}\leq \epsilon$ for all $(t,j)\in\dom\phi$; \emph{ globally pre-attractive} for~\eqref{eq:hybridsystem} if every solution $\phi$ to~\eqref{eq:hybridsystem} is bounded and, if it is complete, then also $\lim_{t+j\to+\infty}\norm[\Amc]{\phi(t,j)}=0$; \emph{globally pre-asymptotically stable} for~\eqref{eq:hybridsystem} if it is both stable and globally pre-attractive. If every maximal solution to~\eqref{eq:hybridsystem} is complete then one may drop the prefix ``pre.''

%% file: traj.tex
\psfrag{001}[][]{\small $\Nmc$}
\psfrag{002}[t][t]{\small $z_1$}
\psfrag{003}[b][b]{\small $z_2$}
\psfrag{004}[tc][tc]{\small $-0.5$}
\psfrag{005}[tc][tc]{\small $0$}
\psfrag{006}[tc][tc]{\small $0.5$}
\psfrag{007}[tc][tc]{\small $1$}
\psfrag{008}[tc][tc]{\small $1.5$}
\psfrag{009}[tc][tc]{\small $2$}
\psfrag{010}[tc][tc]{\small $2.5$}
\psfrag{011}[cr][cr]{\small $-1$}
\psfrag{012}[cr][cr]{\small $-0.5$}
\psfrag{013}[cr][cr]{\small $0$}
\psfrag{014}[cr][cr]{\small $0.5$}
\psfrag{015}[cr][cr]{\small $1$}
\psfrag{016}[cr][cr]{\small $1.5$}
\psfrag{017}[cr][cr]{\small $2$}
\psfrag{018}[cl][cl]{\small Section~\ref{sec:start} with $q(0,0)=-1$}
\psfrag{019}[cl][cl]{\small Section~\ref{sec:start} with $q(0,0)=1$}
\psfrag{020}[cl][cl]{\small Section~\ref{sec:back} with $q(0,0)=-1$}
\psfrag{021}[cl][cl]{\small Section~\ref{sec:back} with $q(0,0)=1$}

%% file: plots.tex
\psfrag{001}[t][t]{\small $t$}
\psfrag{002}[b][b]{\small $\norm{\th(t)-\theta}$}
\psfrag{003}[b][b]{\small $\norm{z(t)}$}
\psfrag{004}[tc][tc]{\small $0$}
\psfrag{005}[tc][tc]{\small $1$}
\psfrag{006}[tc][tc]{\small $2$}
\psfrag{007}[tc][tc]{\small $3$}
\psfrag{008}[tc][tc]{\small $4$}
\psfrag{009}[tc][tc]{\small $5$}
\psfrag{010}[tc][tc]{\small $6$}
\psfrag{011}[tc][tc]{\small $7$}
\psfrag{012}[tc][tc]{\small $8$}
\psfrag{013}[tc][tc]{\small $9$}
\psfrag{014}[tc][tc]{\small $10$}
\psfrag{015}[cr][cr]{\small $0$}
\psfrag{016}[cr][cr]{\small $0.2$}
\psfrag{017}[cr][cr]{\small $0.4$}
\psfrag{018}[cr][cr]{\small $0.6$}
\psfrag{019}[cr][cr]{\small $0.8$}
\psfrag{020}[cr][cr]{\small $1$}
\psfrag{021}[cr][cr]{\small $1.2$}
\psfrag{022}[cr][cr]{\small $0$}
\psfrag{023}[cr][cr]{\small $0.5$}
\psfrag{024}[cr][cr]{\small $1$}
\psfrag{025}[cr][cr]{\small $1.5$}
\psfrag{026}[cr][cr]{\small $2$}
\psfrag{027}[cr][cr]{\small $2.5$}
\psfrag{028}[cl][cl]{\small Section~\ref{sec:start} with $q(0,0)=-1$}
\psfrag{029}[cl][cl]{\small Section~\ref{sec:start} with $q(0,0)=1$}
\psfrag{030}[cl][cl]{\small Section~\ref{sec:back} with $q(0,0)=-1$}
\psfrag{031}[cl][cl]{\small Section~\ref{sec:back} with $q(0,0)=1$}